\documentclass[
showpacs, preprintnumbers,
nofootinbib, aps, prd, superscriptaddress,twocolumn,
10pt, showkeys, notitlepage, longbibliography]{revtex4-1}

\usepackage[dvips]{graphicx}
\usepackage{bm,latexsym,amsmath,amssymb,amsfonts,times}
\usepackage{color}
\usepackage{natbib}
\usepackage{times}
\usepackage{graphicx}
\usepackage{dcolumn}
\usepackage{mathtools}

\def\be {\begin{equation}}
\def\ee {\end{equation}}
\def\ba {\begin{eqnarray}}
\def\ea {\end{eqnarray}}

\def\bi {\begin{itemize}}
\def\ei {\end{itemize}}

\newcommand{\cR}{{\cal R}}

\newcommand\beq{\begin{eqnarray}}
\newcommand\eeq{\end{eqnarray}}

\newcommand\bA{{\bar A}}
\newcommand\bF{{\bar F}}

\usepackage[linktocpage,breaklinks]{hyperref}
\definecolor{venetianred}{rgb}{0.78, 0.03, 0.08}
\definecolor{grey}{rgb}{0.25, 0.25, 0.28}
\definecolor{darkmidnightblue}{rgb}{0.0, 0.2, 0.4}
\definecolor{egyptianblue}{rgb}{0.06, 0.2, 0.65}
\definecolor{darkblue}{rgb}{0.0, 0.0, 0.55}
\hypersetup{colorlinks,breaklinks,
            linkcolor=red,urlcolor=darkblue,
            anchorcolor=red,citecolor=blue}

\def\X5sp{{\rm X}_5}
\def\Y3sp{{\rm Y}_3}
\def\Z3sp{{\rm Z}_3}

\begin{document}

\title{Vector boson star solutions with a quartic order self-interaction}

\author{Masato Minamitsuji}
\email{masato.minamitsuji@ist.utl.pt}
\affiliation{Centro Multidisciplinar de Astrofisica - CENTRA,
Instituto Superior Tecnico - IST,
Universidade de Lisboa - UL,
Avenida Rovisco Pais 1, 1049-001, Portugal.}

\begin{abstract}
We investigate boson star (BS) solutions in the Einstein-Proca theory with the quartic order self-interaction of the vector field $\lambda (A^\mu\bar{A}_\mu)^2/4$ and the mass term $\mu \bar{A}^\mu A_\mu/2$, where $A_\mu$ is the complex vector field and ${\bar A}_\mu$ is the complex conjugate of $A_\mu$, and $\lambda$ and $\mu$ are the coupling constant and the mass of the vector field, respectively. The vector BSs are characterized by the two conserved quantities, the Arnowitt-Deser-Misner (ADM) mass and the Noether charge associated with the global $U(1)$ symmetry. We show that in comparison with the case without the self-interaction $\lambda=0$, the maximal ADM mass and Noether charge increase for $\lambda>0$ and decrease for $\lambda<0$. We also show that there exists the critical central amplitude of the temporal component of the vector field above which there is no vector BS solution, and for $\lambda>0$ it can be expressed by the simple analytic expression. For a sufficiently large positive coupling $\Lambda:=M_{pl}^2\lambda /(8\pi\mu^2) \gg 1$, the maximal ADM mass and Noether charge of the vector BSs are obtained from the critical central amplitude and of ${\cal O}[\sqrt{\lambda}M_{pl}^3/\mu^2 \ln (\lambda M_{pl}^2/\mu^2)]$, which is different from that of the scalar BSs, ${\cal O}(\sqrt{\lambda_\phi}M_{pl}^3/\mu_\phi^2)$, where $\lambda_\phi$ and $\mu_\phi$ are the coupling constant and the mass of the complex scalar field.
\end{abstract}
\date{\today}
\maketitle

\section{Introduction}
\label{sec1}

The recent detection of gravitational waves (GWs) 
from merging black holes (BHs) and neutron stars (NSs)
by the LIGO and Virgo collaborations
\cite{Abbott:2016blz,Abbott:2016nmj,TheLIGOScientific:2017qsa}
has opened new opportunities
to test gravitational and fundamental physics 
in the extremely high density and/or high curvature regions. 
The near-future detection of GWs will be able to test modifications of general relativity (GR)
in strong gravity regimes
in terms of the existence of the hairy BHs \cite{Berti:2015itd,Herdeiro:2015waa}
and the universal relations for NSs \cite{Doneva:2017jop}.

Although the data of LIGO and Virgo 
are highly consistent 
with the theoretical GW waveforms predicted from coalescing BHs and NSs in GR
so far,
they have not excluded the possibility of modified gravity theories
and/or the existence of 
other more exotic compact objects yet,
and the future GW measurements would be able to test them 
more precisely \cite{Berti:2018cxi,Berti:2018vdi,Cardoso:2017cqb}.
One of the candidates of more exotic compact objects is a boson star (BS)
which is a gravitationally bound nontopological solitonic object
in a bosonic field theory. 
If the existence of the BSs could be verified through the near-future GW observations,
it may also give us a direct evidence of extra degrees of freedom 
in modified gravity theories.
The BSs are characterized by 
the two conserved quantities,
namely, 
the Arnowitt-Deser-Misner (ADM) mass $M$
and  
the Noether charge $Q$ associated with the global $U(1)$ symmetry of the field space.
$M$ and $Q$
correspond to the gravitational mass and the number of particles
inside a BS system, respectively,
and a BS is gravitationally bound
if it possesses the positive binding energy, $\mu_\phi Q-M>0$,
where $\mu_\phi$ is the mass of the scalar field.
The BS solutions have been first constructed 
in the Einstein-scalar theory 
with the mass term $\mu_\phi^2 |\phi|^2/2$ \cite{Kaup:1968zz,Ruffini:1969qy,Friedberg:1986tp,Jetzer:1991jr}.
The radial perturbation analysis
about the BS solutions \cite{Gleiser:1988rq,Jetzer:1991jr,Lee:1988av,Gleiser:1988ih}
has revealed that 
the critical solutions dividing the stable BSs and the unstable ones 
correspond to 
those with the maximal ADM mass and Noether charge \cite{Gleiser:1988ih,Hawley:2000dt}.
GW signatures of the binary BSs
would be distinguishable from those of the binary BHs and NSs
as the consequence of
the different tidal deformabilities \cite{Sennett:2017etc}. 

The maximal ADM mass depends on the potential of the scalar field \cite{Schunck:2003kk}. 
In the Einstein-scalar theory 
only with the mass term $\mu_\phi^2 |\phi|^2/2$,
it is of ${\cal O} (M_{pl}^2/\mu_\phi)$,
where $M_{pl}=\sqrt{\hbar c/G}= 1.221\times 10^{19} {\rm GeV}/c^2$ is the Planck mass
(in the rest we work in the units of $c=\hbar=1$),
which is much smaller than the Chandrasekhar mass
for fermions with the mass $\mu_\phi$,
of ${\cal O} (M_{pl}^3/\mu_\phi^2)$,
by assuming that $\mu_\phi\ll M_{pl}$.
In the theory with 
both the quartic order self-interaction $\lambda_\phi |\phi|^4/4$
as well as
the mass term $\lambda_\phi |\phi|^2/2$,
it becomes of the same order as the Chandrasekhar mass 
for the ferminons with the mass $\mu_\phi$,
of ${\cal O} (\sqrt{\lambda_\phi}M_{pl}^3/\mu_\phi^2)$ \cite{Colpi:1986ye}.

The BS solutions exist not only in the Einstein-scalar theory 
but also in the Einstein-Proca theory \cite{Brito:2015pxa,Garcia:2016ldc,Brihaye:2017inn}.
In the Einstein-Proca theory with the mass term $\mu^2 \bar{A}^\mu A_\mu/2$,
where $A_\mu$ is the vector field and 
$\bar{A}_\mu$ is the complex conjugate of $A_\mu$, 
the properties of the vector BSs are quite similar 
to those of the BSs in the Einstein-scalar theory with the mass term \cite{Brito:2015pxa}.
The critical solution dividing the stable and unstable vector BSs
corresponds to the solution with the maximal ADM mass and Noether charge,
and the maximal ADM mass of a vector BS is of ${\cal O}(M_{pl}^2/\mu)$.

The question which we addreess in this paper
is
how the maximal ADM mass and Noether charge of the BS solutions
in the Einstein-Proca theory
with the quartic order self-interaction of the vector field
$\lambda (\bar{A}^\mu A_{\mu})^2/4$ 
as well as the mass term $\mu^2 \bar{A}^\mu A_{\mu}/2$
are related to the mass and coupling constant of the vector field, $\mu$ and $\lambda$,
and 
whether they are similar to those 
in the Einstein-scalar theory
with the quartic order self-interaction.
The solutions of nontopological solitions 
in the complex vector field theories with the nonlinear self-interaction potential
in the flat and curved spacetimes
have been presented in Refs.  \cite{Loginov:2015rya,Brihaye:2016pld,Brihaye:2017inn}.
While the results presented in this paper 
have some overlap with those presented in Ref. \cite{Brihaye:2017inn},
we focus more on the role of the quartic order self-interaction 
in the self-gravitating vector BS backgrounds,
and derive the quantitative dependence 
of the physical properties of the BSs on $\lambda$.

The paper is constructed as follows:
In Sec. \ref{sec2}, 
we introduce the Einstein-Proca theory with the mass and 
the quartic order self-interaction,
and derive a set of equations to find the structure of the vector BSs.
In Sec. \ref{sec3}, we numerically construct the vector BS solutions 
and make arguments about their properties.
The last Sec. \ref{sec4} is devoted to giving a brief summary and conclusion.

\section{Vector boson star solutions}
\label{sec2}

\subsection{Theory}
\label{sec2a}

We consider the Einstein-Proca theory with the quartic order self-interaction
as well as the mass term;
\begin{align}
\label{action}
S&=\int d^4x \sqrt{-g}
\left[
 \frac{1}{2\kappa^2}R
-\frac{1}{4} F^{\mu\nu}{\bar F}_{\mu\nu}
\right.
\nonumber \\
&
\left.
-\frac{1}{2} \mu^2 A^\mu {\bar A}_\mu
-\frac{1}{4}\lambda \left(A^\mu {\bar A}_\mu\right)^2
\right],
\end{align}
where 
the greek indices $(\mu,\nu,...)$ run the four-dimensional spacetime,
$g_{\mu\nu}$ is the metric tensor,
$g^{\mu\nu}:=\left(g_{\mu\nu}\right)^{-1}$,
$g={\rm det} (g_{\mu\nu})$ is the determinant of $g_{\mu\nu}$,
$R$ is the scalar curvature associated with $g_{\mu\nu}$,
$A_\mu$ is the complex vector field
and $\bar{A}_\mu$ is the complex conjugate of $A_\mu$,
$F_{\mu\nu}:= \partial_\mu A_\nu-\partial_\nu A_\mu$
is the field strength
and 
$\bar{F}_{\mu\nu}:= \partial_\mu \bar{A}_\nu-\partial_\nu \bar{A}_\mu$
is its complex conjugate,
$\kappa^2:= 8\pi G$ with $G$ being Newton's constant, 
$\mu$ is the mass of the complex vector field,
and 
the dimensionless coupling constant
$\lambda$ measures the strength of the quartic order self-interaction of the vector field.
Throughout the paper we set $c=\hbar=1$, and 
in these units
the Planck mass $M_{pl}$ is given by $M_{pl}=1/\sqrt{G}$.
Because of the different sign convention,
$\lambda$ in this paper corresponds 
to $(-\lambda)$ in Refs. \cite{Brihaye:2016pld,Brihaye:2017inn}.

Varying the action \eqref{action} with respect to $g_{\mu\nu}$,
we obtain the gravitational field equations of motion (EOM)
\begin{align}
\label{grav_eq}
0={\cal E}_{\mu\nu}
&:= T^{(A)}_{\mu\nu}-\frac{1}{\kappa^2}G_{\mu\nu},
\end{align}
where
\begin{align}
\label{tf}
T^{(A)}_{\mu\nu}
&:=
 \frac{1}{2}
\left( F_{\mu\rho} {\bar F}_\nu{}^\rho
+ {\bar F}_{\mu\rho} F_\nu{}^\rho
\right) 
-\frac{1}{4} g_{\mu\nu}F^{\rho\sigma} {\bar F}_{\rho\sigma}
\nonumber\\
&
+
\frac{\mu^2}{2}
\left[
   A_\mu {\bar A}_\nu
+ A_\nu {\bar A}_\mu
-
g_{\mu\nu} \bar{A}^\rho A_\rho
\right]
\nonumber\\
&+
\frac{\lambda}{2}
\left[
\left(\bar{A}^\rho  A_\rho\right)
 \left( 
A_\mu {\bar A}_\nu
+ A_\nu {\bar A}_\mu
\right)
-\frac{1}{2}
g_{\mu\nu}
\left(\bar{A}^\rho A_\rho\right)^2
\right],
\end{align}
represents the energy-momentum tensor of the vector field.
Similarly, 
varying the action \eqref{action} with respect to
$A_\mu$ and $\bar{A}_\mu$,
we obtain the EOM for the vector field  
\begin{align}
\label{vector_eq}
0= {\cal F}_\nu
&:=
 \nabla^\mu F_{\mu\nu}
-\left[
 \mu^2
+\lambda \left(\bar{A}^\rho  A_\rho\right)
\right]
A_\nu,
\end{align}
and its complex conjugate $\overline{\cal F}_\nu=0$,
respectively.
Acting the derivative $\nabla^\nu$ on Eq. \eqref{vector_eq},
we obtain the constraint relation
\begin{align}
\label{gauge}
0={\cal G}&:=
\nabla^\nu
\left\{
\left[
 \mu^2
+\lambda \left(\bar{A}^\rho  A_\rho\right)
\right]
A_\nu
\right\}.
\end{align}
Similarly, $\nabla^\nu\overline{\cal F}_\nu=:\bar{\cal G}=0$.
Note that 
in the theory \eqref{action}
there is the global $U(1)$ symmetry,
namely the symmetry 
under the transformation $A_\mu\to e^{i \alpha} A_\mu$ where $\alpha$ is a constant,
and the associated Noether current
is given by 
\begin{align}
\label{noether}
j^\mu=\frac{i}{2}
\left(
 \bF^{\mu\nu} A_\nu
-F^{\mu\nu} \bA_\nu
\right),
\end{align}
which satisfies the conservation law, $\nabla_\mu  j^\mu=0$.

\subsection{Static and spherically symmetric spacetime}
\label{sec2b}

We consider a static and spherically symmetric spacetime
\begin{align}
\label{metric_ansatz}
g_{\mu\nu}dx^\mu dx^\nu
&=-\sigma(r)^2 
\left(1-\frac{2m(r)}{r}\right) d{\hat t}^2 
\nonumber\\
&+ 
   \left(1-\frac{2m(r)}{r}\right)^{-1}
dr^2
 + r^2 d\Omega_2^2,
\end{align}
where $\hat t$ and $r$ are the time and radial coordinates, 
$d\Omega_2^2$ is the metric of the unit two sphere,
and 
$m(r)$ and $\sigma(r)$ 
depend only on the radial coordinate $r$.
Correspondingly, 
we consider the ansatz for the vector field \cite{Brito:2015pxa}
\begin{align}
\label{vector_ansatz}
A_\mu dx^\mu
&=e^{-i{\hat\omega} {\hat t}}
 \left(
     a_0(r)d{\hat t}
   +i a_1 (r)dr
 \right),
\end{align}
where $a_0(r)$ and $a_1(r)$ depend only on $r$.
For the vector BS solutions, 
the frequency ${\hat\omega}$ is assumed to be real and positive,
such that the vector field neither grows nor decays in time.
The ansatz \eqref{vector_ansatz}
is compatible with the static and spherically symmetric spacetime \eqref{metric_ansatz},
as it does not give rise to the explicit time dependence of the energy-momentum tensor.
In order to find $m$, $\sigma$, $a_0$, and $a_1$ numerically,
we rewrite EOMs \eqref{grav_eq}, \eqref{vector_eq} and \eqref{gauge}
into a set of the evolution equations
with respect to $r$.

\subsection{Evolution equations}
\label{sec2c}

From our ansatz \eqref{metric_ansatz} and \eqref{vector_ansatz},
the nontrivial components of the gravitational EOMs \eqref{grav_eq}
are given by 
\begin{align}
\label{grav_eq0}
&
{\cal E}^{\hat t}{}_{\hat t}=0,
\quad 
{\cal E}^r{}_r=0,
\quad
{\cal E}^i{}_i=0,
\end{align} 
where the indices $(i,j,\cdots)$ run the directions of the two sphere.
Similarly,
the nontrivial components of the vector field EOM \eqref{vector_eq}
are given by 
\begin{align}
\label{vector_eq0}
&{\cal F}_{\hat t}=0,
\quad
{\cal F}_r=0.
\end{align}
The complex conjugates of Eq. \eqref{vector_eq0}
give the same equations as Eq. \eqref{vector_eq0}
and need not be considered separately.
All these equations are related by 
\begin{align}
\label{bianchi}
\nabla_\mu 
{\cal E}^\mu{}_r
&=
-\frac{1}{2}
  g^{{\hat t}{\hat t}} 
 \left(
{\cal F}_{\hat t} 
\times 
\bar{F}_{{\hat t}r}
 +
\bar{\cal F}_{\hat t}
\times 
F_{{\hat t}r}\right) 
\nonumber\\
&+\frac{1}{2}
\left(
{\cal G} \bA_r
+
 {\bar{\cal  G}} A_r
\right).
\end{align}

First, ${\cal E}^{\hat t}{}_{\hat t}=0$ and ${\cal E}^r{}_r=0$ 
can be arranged to give the evolution equations for $m$ and $\sigma$
in the $r$ direction  as
\begin{subequations}
\label{ms_evolv}
\begin{align}
m'&= F_m\left[a_0,a_0',a_1,m,\sigma \right],
\\
\sigma'&= F_\sigma\left[a_0,a_1,m,\sigma\right],
\end{align}
\end{subequations}
where a prime denotes the derivative with respect to $r$,
and $F_m$ and $F_\sigma$
are the nonlinear combinations of the given variables as
\begin{widetext}
\begin{subequations}
\label{fmfsigma}
\begin{align}
F_m
&:=\frac{\kappa^2}{8(r-2m)^2\sigma^4}
\left[
-3r^4\lambda a_0^4
+2r^2 a_0^2 \left(r\mu^2 +\lambda a_1^2 (r-2m)\right)
 (r-2m) \sigma^2
+\lambda a_1^4 (r-2m)^4\sigma^4
\right.
\nonumber\\
&
\left.
+2r a_1^2 (r-2m)^2 \sigma^2 
\left(r{\hat \omega}^2+\mu^2 (r-2m)\sigma^2\right)
-4r^2 {\hat\omega} a_1 (r-2m)^2\sigma^2a_0'
+2r^2 (r-2m)^2\sigma^2a_0'^2
\right],
\\
F_\sigma
&:= 
\frac{\kappa^2}{2 (r-2m)^3\sigma^3}
\left[
-r^2\lambda a_0^2 
+
\left(
r\mu^2 +\lambda a_1^2 (r-2m)
\right)
(r-2m)
\sigma^2
\right)
\left(
r^2a_0^2 +a_1^2 (r-2m)^2\sigma^2
\right].
\end{align}
\end{subequations}
\end{widetext}
Similarly,
${\cal F}_{\hat t}=0$ and  ${\cal G}=0$ 
can be arranged to be 
\begin{subequations}
\label{a0a1_evolv}
\begin{align}
a_0''&= {\cal H}^{-1}F_0\left[a_0,a_0',a_1,m,m',\sigma,\sigma'\right],
\\
a_1'&= {\cal H}^{-1} F_1\left[a_0,a_0',a_1,m,m',\sigma,\sigma' \right],
\end{align}
\end{subequations}
where $F_0$ and $F_1$ are regular combinations of the given variables
which are too involved to be shown explicitly, 
and 
\begin{align}
\label{h}
{\cal H}
:=
-r^2 \lambda a_0^2
+
\left(
r\mu^2 +3\lambda a_1^2 \left(r-2m\right)
\right)
 \left(r-2m\right)
\sigma^2.
\end{align}
Substituting Eqs. \eqref{ms_evolv} into the left-hand side of Eqs. \eqref{a0a1_evolv}
and eliminating $m'$ and $\sigma'$,
we obtain 
\begin{subequations}
\label{a0a1_evolv2}
\begin{align}
a_0''&={\cal H}^{-1}
 \tilde{F}_0\left[a_0,a_0',a_1,m,\sigma\right],
\\
a_1'&={\cal H}^{-1}
 \tilde{F}_1 \left[a_0,a_0',a_1,m,\sigma \right],
\end{align}
\end{subequations}
where ${\tilde F}_0$ and $\tilde F_1$ are regular combinations of the given variables
which are also too involved to be shown explicitly.

Equations \eqref{ms_evolv} and \eqref{a0a1_evolv2}
are integrated from a point sufficiently close to $r=0$
with the boundary conditions given in Sec. \ref{sec2d}.
For a given set of parameters,
if ${\cal H}$ vanishes at some finite radius,
Eqs. \eqref{ms_evolv} and \eqref{a0a1_evolv2}
cannot be integrated further beyond this point,
and then no vector BS solution exists.

\subsection{Boundary conditions}
\label{sec2d}

Solving Eqs. \eqref{ms_evolv} and \eqref{a0a1_evolv2}
in the vicinity of $r=0$,
we find
\begin{subequations}
\label{bc}
\begin{align}
a_0
&= f_0 
- \frac{f_0}{6\sigma_0^2}
\left(
 f_0^2\lambda 
-\mu^2\sigma_0^2
+{\hat \omega}^2
\right)
  r^2
+{\cal O} \left(r^4\right),
\\
a_1
&=-
\frac{f_0 {\hat\omega}}{3\sigma_0^2}
r
+{\cal O} \left(r^3\right),
\\
m
&=
\frac{\kappa^2f_0^2}{24\sigma_0^4}
\left(
-3f_0^2\lambda
+2 \mu^2 \sigma_0^2
\right)
r^3
+{\cal O} \left(r^5\right),
\\
\sigma
&=\sigma_0
+
\frac{\kappa^2 f_0^2}{4\sigma_0^3}
\left(
-f_0^2 \lambda +\mu^2 \sigma_0^2
\right)
r^2
+{\cal O} \left(r^4\right).
\end{align}
\end{subequations}
Equation \eqref{bc} evaluated at $r=r_1$ which is sufficiently close to $r=0$
gives the boundary conditions 
to integrate Eqs. \eqref{ms_evolv} and \eqref{a0a1_evolv2} 
numerically
from $r=r_1$
to a sufficiently large value of $r$.

For $\lambda=0$, 
${\cal H}$ defined in Eq. \eqref{h}
never crosses $0$
and the vector BS solutions exist
for an arbitrary value of 
the central amplitude of the temporal component of the vector field, $f_0>0$.
For $\lambda\neq 0$,
expanding ${\cal H}$ in the vicinity of $r=0$
with Eq. \eqref{bc},
we obtain 
\begin{align}
\label{calh}
{\cal H}
=
\left(
-f_0^2\lambda +\mu^2 a_0^2
\right)
r^2
+
{\cal O} (r^4).
\end{align}
For $\lambda>0$,
even if ${\cal H}<0$ in the vicinity of $r=0$,
as the ${\cal O}(r^4)$ terms in Eq. \eqref{calh} become important,
${\cal H}$ starts to increase and cross $0$.
Thus, a regular BS solution can be obtained only for
\begin{align}
\label{bound_amp}
0<f_0\leq f_{0,{\rm crit}}:= \frac{\mu\sigma_0}{\sqrt{\lambda}},
\end{align}
which was numerically confirmed,
where $f_{0,{\rm crit}}$ corresponds to the critical central amplitude
of the temporal component of the vector field.
On the other hand,
for $\lambda<0$,
even if ${\cal H}>0$ in the vicinity of $r=0$,
the higher order corrections to Eq. \eqref{calh}
make ${\cal H}$ decrease and cross $0$.
Thus, also for $\lambda<0$,
we numerically confirmed the existence of the critical central amplitude 
of the temporal component of the vector field $f_{0,{\rm crit}}$,
although it cannot be expressed analytically.

For ${\hat\omega}$ chosen to be the correct lowest eigenvalue
of the vector BS 
for a given set of parameters,
$m$ and $\sigma$ exponentially approach constant values, 
$m_\infty>0$ and $\sigma_\infty>0$,
respectively,
while
$a_0$ and $a_1$ exponentially approach $0$ as $e^{-\sqrt{\mu^2-{\hat\omega}^2/\sigma_\infty^2}r}$.
Thus,
the metric exponentially approaches the Schwarzschild form
\begin{align}
\label{metric}
ds^2
&\to 
-\sigma_\infty^2 
  \left(
   1-\frac{2m_\infty}{r}
  \right)d{\hat t}^2
\nonumber\\ 
&+  
\left(1-\frac{2m_\infty}{r}\right)^{-1}
dr^2
+r^2d\Omega_2^2,
\end{align}
where the proper time measured 
by the observer at $r=\infty$ is given by 
$t=\sigma_\infty {\hat t} $,
and 
correspondingly 
the proper frequency $\omega$ is given by
\begin{align}
\label{hatom}
{\omega}:=\frac{{\hat\omega}}{\sigma_\infty}.
\end{align}
The condition for the exponential fall-off 
properties $e^{-\sqrt{\mu^2-\omega^2}r}$
requires $\omega <\mu$.
In the limit $f_0\to 0$, 
$\omega\to \mu$,
and the Minkowski solution is obtained.

By the rescalings of 
\begin{align}
\label{dimless}
&
{\hat\omega}\to \frac{{\hat\omega}}{\mu},
\quad
r\to r \mu,
\quad
m \to \mu m,
\quad
\sigma \to \sigma,
\nonumber\\
&
a_0 \to \kappa a_0,
\quad
a_1\to \kappa a_1,
\quad
\lambda 
\to 
\frac{\lambda}
{\mu^2\kappa^2},
\end{align} 
the evolution equations \eqref{ms_evolv} and \eqref{a0a1_evolv2} 
can be rewritten into the form without $\mu$ and $\kappa$,
and in the rescaled equations the strength of the self-interaction is measured by 
the dimensionless coupling constant
\begin{align}
\label{Lambda}
\Lambda :=
\frac{\lambda}
{\mu^2\kappa^2}
=\frac{M_{pl}^2}{(8\pi)\mu^2}  \lambda.
\end{align}
For the numerical analysis,
we may set $\mu=\kappa=1$ and 
as the result $\lambda=\Lambda$,
as it is  straightforward 
to give back the dependence
of the physical quantities on $\mu$ and $\kappa$,
once the vector BS solutions are numerically obtained for $\mu=\kappa=1$.
In addition, 
as $\sigma_0$ corresponds to the degree of freedom of the time rescaling
without loss of generality,
we may also set $\sigma_0=1$.
Thus,
only the remaining physical parameters are $f_0$ and $\Lambda$.

\begin{figure}[h]
\unitlength=1.1mm
\begin{center}
  \includegraphics[height=5.0cm,angle=0]{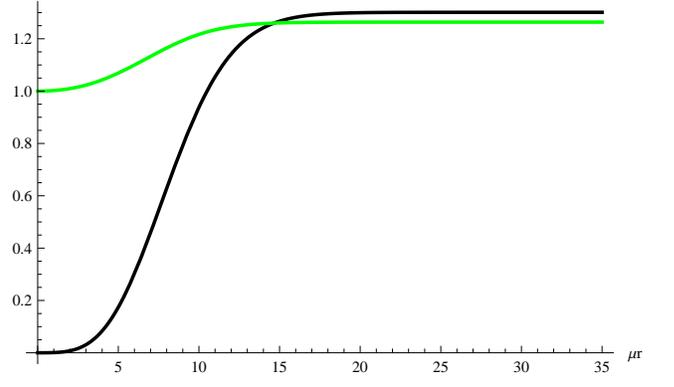}
\caption{
$\mu m$ and $\sigma$ are shown as the functions of 
$\mu r$ for $\Lambda=10$ and $f_0=1/(10\kappa)$.
The black and green curves correspond to 
$\mu m$ and $\sigma$, respectively.
}
  \label{figms}
\end{center}
\end{figure}

\begin{figure}[h]
\unitlength=1.1mm
\begin{center}
  \includegraphics[height=5.0cm,angle=0]{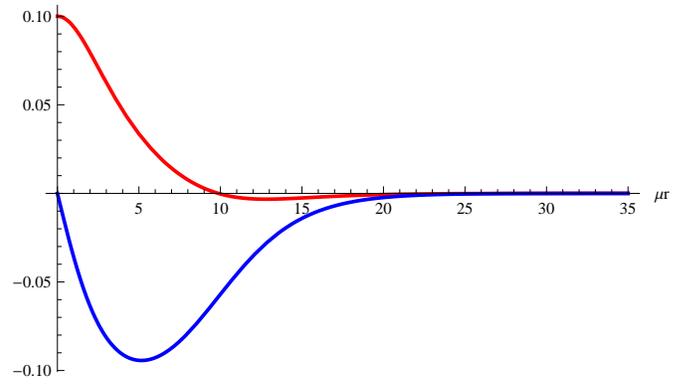}
\caption{
$\kappa a_0$ and $\kappa a_1$ are shown as the functions of 
$\mu r$ 
for $\Lambda=10$ and $f_0=1/(10\kappa)$.
The red and blue curves correspond to 
$\kappa a_0$ and $\kappa a_1$, respectively.
}
  \label{figa0a1}
\end{center}
\end{figure}

In Figs. \ref{figms} and \ref{figa0a1},
$(m,\sigma)$ and $(a_0, a_1)$
are shown as the functions of $r$
for $\Lambda=10$ and $f_0=1/(10\kappa)$, respectively.
Note that 
the dimensionless quantities introduced in \eqref{dimless}
are different from those introduced in Ref. \cite{Brihaye:2017inn}
where in the dimensionless equations of motion
$a_0$ and $a_1$ are rescaled 
to the dimensionless quantities,
$(\sqrt{|\lambda|}/\mu) a_0$ and $(\sqrt{|\lambda|} /\mu) a_1$,
respectively.
While Ref. \cite{Brihaye:2017inn} focused on 
the importance of the coupling to gravity
measured by the dimensionless parameter 
$\alpha:=(\kappa^2\mu^2)/\lambda[=1/\Lambda]$
on the nontopological soliton backgrounds in the flat spacetime,
we focus more on the role of the quartic order self-interaction of the  
vector field
on the self-gravitating vector BS backgrounds,
and derive the quantitative dependence 
of the physical properties of BSs on the coupling constant $\lambda$.
Thus, 
although there are similarities of our results to those in Ref. \cite{Brihaye:2017inn},
we make arguments about the properties of the vector BSs from the different perspectives.
In contrast to the case of the scalar BSs,
in the case of the vector BSs 
$a_0$ has a single node
before approaching $0$.

\subsection{ADM mass, Noether charge, and binding energy}
\label{sec2e}

We then evaluate the conserved quantities characterizing the vector BSs.
The first is the ADM mass 
\begin{align}
\label{adm}
M:=\frac{m_\infty}{G}=M_{pl}^2 m_\infty,
\end{align}
which is associated with the time translational symmetry.
The second is 
the Noether charge associated with the global $U(1)$ symmetry,
which is given by 
integrating $j^{\hat t}$ in Eq. \eqref{noether}
over a constant-${\hat t}$ hypersurface
\begin{align}
\label{charge}
Q
=\int_\Sigma d^3x \sqrt{-g} j^{\hat t}
= 4\pi \int_0^\infty 
dr
\frac{
r^2a_1
\left(
 {\hat\omega} a_1
-a_0'
\right)}
{\sigma}.
\end{align}
A BS is gravitationally bound when $\mu Q-M>0$,
and we then define the relative binding energy 
\begin{align}
\label{binding}
b:=\frac{B}{M}=\frac{\mu Q}{M}-1.
\end{align}

In order to discriminate various compact objects
and classify their physical properties,
it is important to define the effective compactness of the vector BS
\begin{align}
\label{compactness}
{\cal C}:=\frac{GM}{\cR}
    =\frac{m_\infty}{\cR},
\end{align}
where the effective radius
of it is given by \cite{Schunck:2003kk}
\begin{align}
\label{radius}
\cR
&
:=
\frac{1}{Q}
\int_\Sigma d^3x 
\sqrt{-g}
 \left(r j^{\hat t}\right)
= 
\frac{4\pi}{Q}
 \int_0^\infty 
dr
\frac{
r^3a_1
\left(
 {\hat\omega} a_1
-a_0'
\right)}
{\sigma}.
\end{align}

\subsection{Comparison with the Einstein-scalar theory}
\label{sec2f}

Before proceeding to the physical properties of the vector BSs,
we briefly review the case of 
the Einstein-scalar theory with the quartic order self-interaction
\begin{align}
\label{action2}
S&=\int d^4x \sqrt{-g}
\left[
 \frac{1}{2\kappa^2}R
-\frac{1}{2} g^{\mu\nu}\partial_\mu \phi \partial_\nu \bar{\phi}
\right.
\nonumber\\
&\left.
-\frac{1}{2} \mu_\phi^2 |\phi|^2
-\frac{1}{4}\lambda_\phi |\phi|^4
\right],
\end{align}
where
$\mu_\phi$ and $\lambda_\phi$ are 
the mass and coupling constant of the complex scalar field, respectively.
In the case without the self-interaction, $\lambda_\phi=0$,
the maximal ADM mass is given 
as \cite{Jetzer:1991jr,Schunck:2003kk},
\begin{align}
M_{\rm max}
\simeq 0.633 \frac{M_{pl}^2}{\mu_\phi},
\end{align}
which is much smaller than
the Chandrasekhar mass for fermions with the same mass.
On the other hand, 
in the case of $\lambda_\phi \left(M_{pl}/\mu_\phi\right)^2 \gg 1$,
no upper bound on the central amplitude of the scalar field  exists,
and the maximal ADM mass is given as
\cite{Colpi:1986ye},
\begin{align}
\label{mmax} 
M_{\rm max}
\simeq 0.062\sqrt{\lambda_\phi}\frac{M_{pl}^3}{\mu_\phi^2}.
\end{align}
Moreover, 
as shown in Fig. \ref{figa0a1}
the vector BS solution
has a single node in $a_0$,
while the scalar BS solution has no node in $|\phi|$.
In Sec. \ref{sec3}
we obtain 
the corresponding relation to Eq. \eqref{mmax}
for the vector BSs
in the presence of the self-interaction $\lambda (\bar{A}^\mu A_\mu)^2/4$.

\section{Properties of Vector Boson Stars}
\label{sec3}

In this section, we make arguments about the properties 
of the vector BS solutions in the theory \eqref{action}
obtained numerically.

\subsection{ADM mass and Noether charge}
\label{sec3a}

\begin{figure}[h]
\unitlength=1.1mm
\begin{center}
  \includegraphics[height=5.0cm,angle=0]{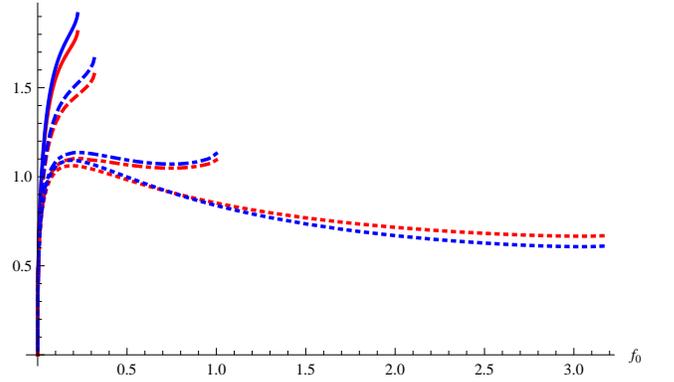}
\caption{
$M$ and $\mu Q$
are shown as the functions of $f_0$ for $\Lambda>0$.
The red and blue curves
correspond to $M$ and $\mu Q$, respectively.
The solid, dashed, dot-dashed, and dotted curves
correspond to 
$\Lambda=20, 10, 1, 0.1$, respectively.
$(M, \mu Q$) and $f_0$ are shown in 
$M_{pl}^2/\mu$ and $M_{pl}/\sqrt{8\pi}[=1/\kappa]$,
respectively.
}
  \label{figmqf}
\end{center}
\end{figure}

In Fig. \ref{figmqf},
the ADM mass $M$ and 
the Noether charge multiplied by $\mu$, $\mu Q$,
are shown as the functions of $f_0$
for several values of $\Lambda\geq 0$ defined in Eq. \eqref{Lambda}.
We numerically confirmed 
that there exists the critical central amplitude of 
the temporal component of the vector field,
$f_{0,{\rm crit}}$,
and 
that it agrees with Eq.  \eqref{bound_amp}.
The behavior for $0<\Lambda\lesssim 1$
is qualitatively similar to that for $\Lambda=0$
except for the existence of $f_{0,{\rm crit}}$;
namely
$M$ and $Q$
take the local maximal values at some intermediate value of $f_0$.
For $\Lambda>1$,
$M$ and $Q$ 
monotonically increase for increasing values of $f_0$,
and their maximal values 
are obtained from the limit of $f_0\to f_{0, \rm crit}$.
On the other hand, 
although we do not show the plots of $M$ and $\mu Q$
as the functions of $f_0$
for $\Lambda<0$,
the values of $M$ and $Q$
become smaller
than those for $\Lambda=0$
for the same values of $f_0$.
For any value of $\Lambda<0$,
we numerically confirmed that 
there is also the critical amplitude of the temporal component of the vector field $f_{0,\rm crit}$,
which cannot be expressed analytically
and becomes smaller for larger $|\Lambda|$.
The maximal values of $M$ and $Q$
correspond to their local maximal values
obtained at the intermediate value of $f_0<f_{0,{\rm crit}}$.
Note that
for all the values of $\Lambda$,
$M=Q=0$ and  
the Minkowski solution with 
the vanishing vector field $A_\mu=0$
is obtained for $f_0=0$.

\begin{figure}[h]
\unitlength=1.1mm
\begin{center}
  \includegraphics[height=5.0cm,angle=0]{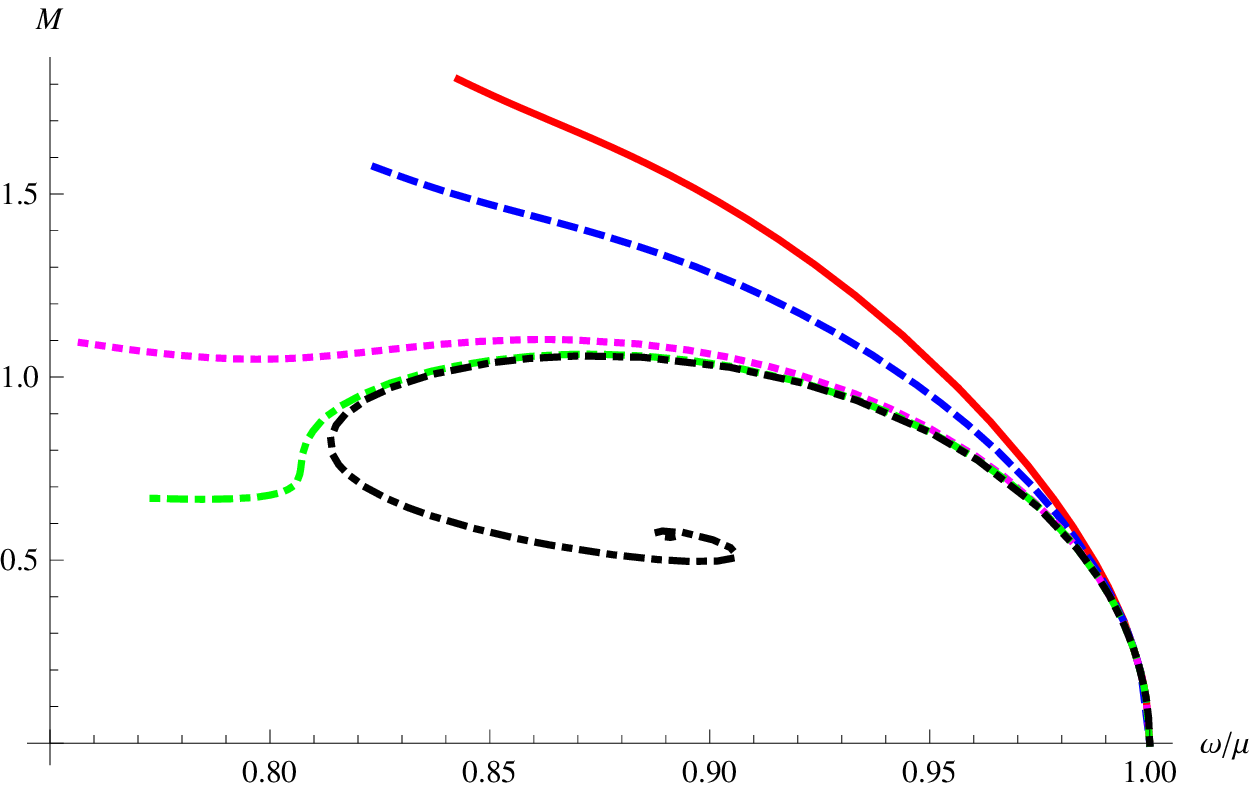}
\caption{
$M$ is shown as the function of $\omega/\mu$
for $\Lambda\geq0$.
The red, blue, magenta, green, and black curves
correspond to the cases of 
$\Lambda=20, 10, 1, 0.1, 0$, respectively.
$M$ is shown in $M_{pl}^2/\mu$.
}
  \label{figmo0}
\end{center}
\end{figure}

In Fig. \ref{figmo0},
$M$ is shown as the function of $\omega$ defined in Eq. \eqref{hatom}
for $\Lambda>0$.
While for $\Lambda=0$
the well-known spiraling behavior is observed 
as shown in Ref. \cite{Brito:2015pxa}
and hence $M$ is the multivalued function of $\omega$,
$M$ eventually increases and becomes the single-valued function of $\omega$
for $0.1 \lesssim \Lambda\lesssim 1$.
Furthermore,
for $\Lambda\gtrsim 1$,
$M$ becomes the monotonically decreasing function of $\omega$.
Note that a quantitatively very similar behavior is obtained for $\mu Q$.

\begin{figure}[h]
\unitlength=1.1mm
\begin{center}
  \includegraphics[height=5.0cm,angle=0]{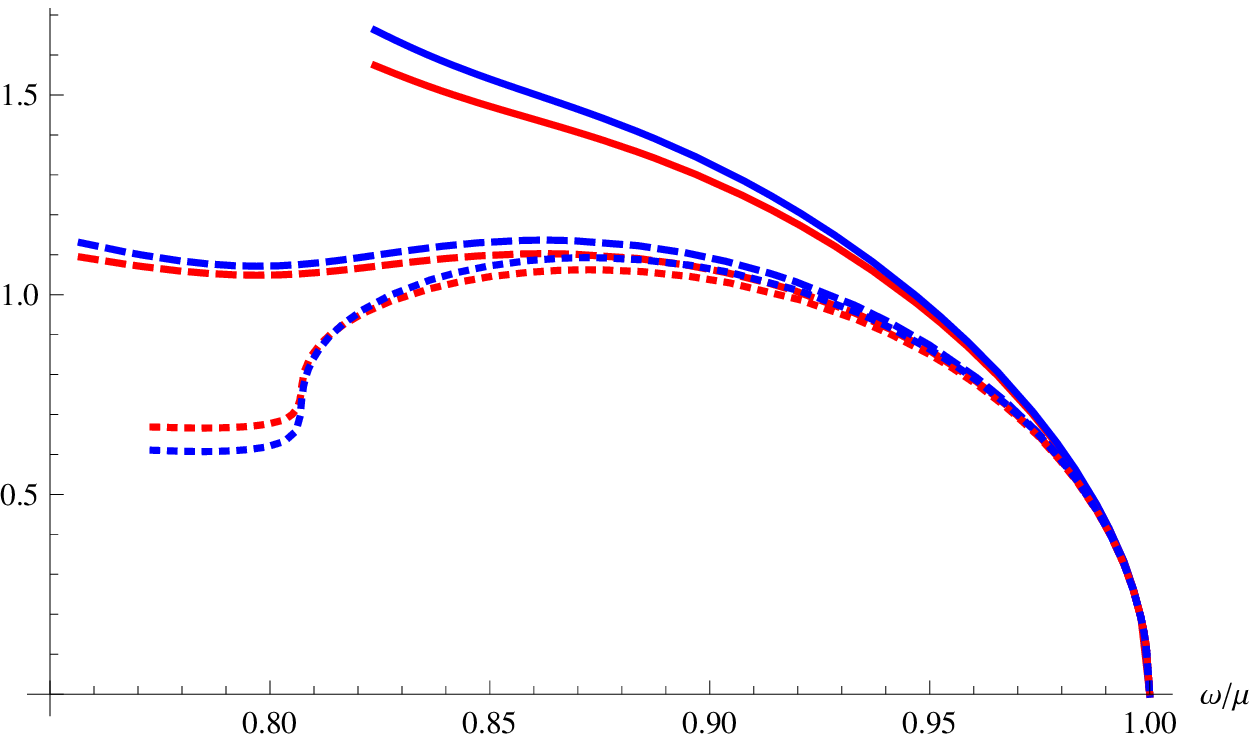}
\caption{
$M$ and $\mu Q$
are compared as the functions of $\omega/\mu$
for $\Lambda>0$.
The red and blue curves
correspond to $M$ and $\mu Q$, respectively.
The solid, dashed, and dotted curves
correspond to 
$\Lambda=10, 1, 0.1$, respectively.
$M$ and $\mu Q$ are shown in 
$M_{pl}^2/\mu$.
}
  \label{figmqo}
\end{center}
\end{figure}

In Fig. \ref{figmqo},
$M$ and $\mu Q$
are compared for $\Lambda=10,1,0.1$, respectively.
For all the cases,
the solutions with the maximal values of $M$ and $Q$
satisfy $M<\mu Q$, namely, $b>0$ in Eq. \eqref{binding},
and these vector BSs are gravitationally bound. 
For $\Lambda \lesssim 0.1$,
$b<0$ for the smaller values of $\omega$
which are obtained from
the values of $f_0$ very close to 
the critical value $f_{0,\rm crit}$, $f_0\lesssim f_{0,{\rm crit}}$.
For $\Lambda\approx 1$, 
the values of $M$ and $\mu Q$ obtained from the limit of $f_0\to f_{0,{\rm crit}}$
become comparable to the local maximal values of $M$ and $\mu Q$
obtained at the intermediate value of $f_0<f_{0,{\rm crit}}$,
respectively.
For $\Lambda\gtrsim 1$,
we obtain
$b>0$ for all values of $\omega$.

\begin{figure}[h]
\unitlength=1.1mm
\begin{center}
  \includegraphics[height=5.0cm,angle=0]{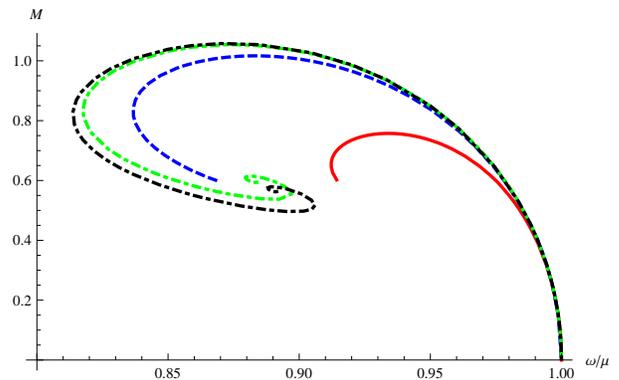}
\caption{
$M$ is shown as the function of $\omega/\mu$
for $\Lambda\leq 0$.
The red, blue, green, and black curves
correspond to the cases of 
$\Lambda=-10, -1, -0.1, 0$, respectively.
$M$ is shown in $M_{pl}^2/\mu$.
}
  \label{figmom}
\end{center}
\end{figure}

\begin{figure}[h]
\unitlength=1.1mm
\begin{center}
  \includegraphics[height=5.0cm,angle=0]{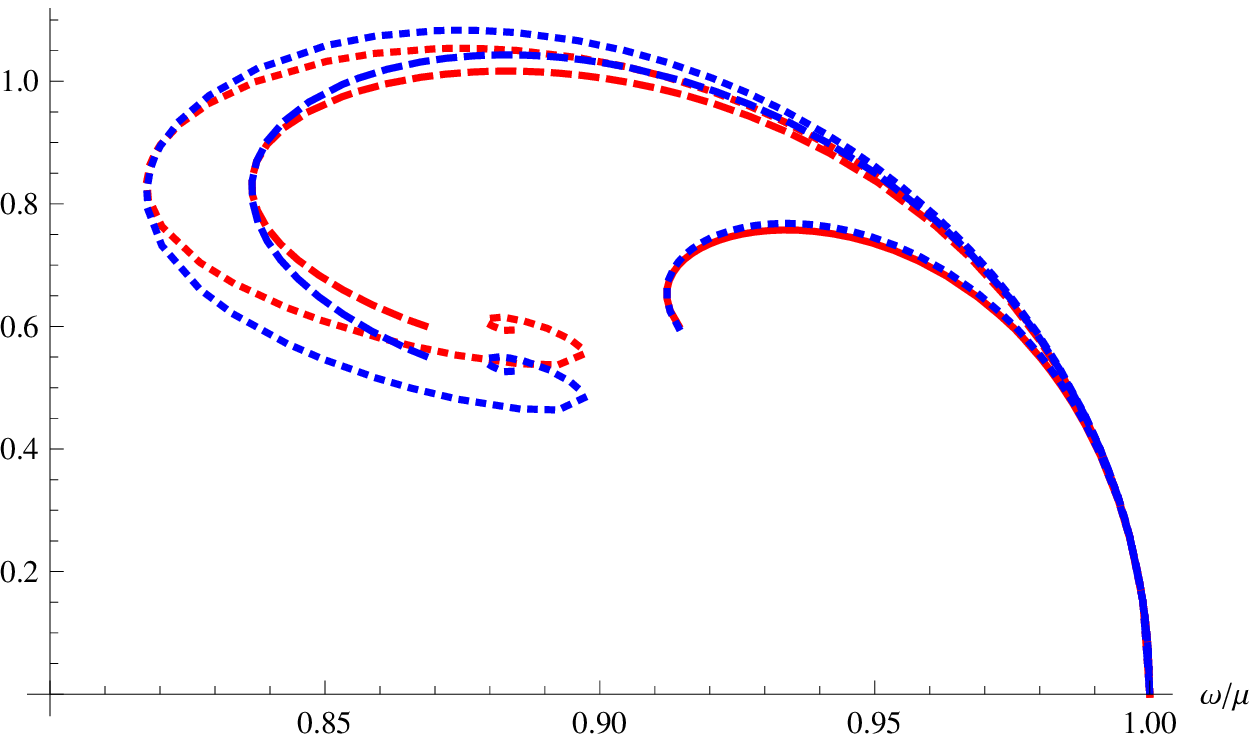}
\caption{
$M$ and $\mu Q$
are compared as the functions of $\omega/\mu$
for $\Lambda<0$.
The red and blue curves
correspond to $M$ and $\mu Q$, respectively.
The solid, dashed, and dotted curves
correspond to 
$\Lambda=-10, -1,-0.1$, respectively.
$M$ and $\mu Q$ are shown in 
$M_{pl}^2/\mu$.
}
  \label{figmqom}
\end{center}
\end{figure}

In Fig. \ref{figmom},
$M$ is shown as the function of $\omega$
for $\Lambda\leq 0$.
As $|\Lambda|$ increases,
$M$ decreases for a fixed value of $f_0$.
Because of the existence of the critical central amplitude 
of the temporal component of the vector field $f_{0,\rm crit}$,
for larger values of $|\Lambda|$
the spiraling behavior is eventually resolved.
The maximal value of $M$ is obtained for $f_0<f_{0,\rm crit}$.
The quantitatively similar behavior is obtained for $\mu Q$.
In Fig. \ref{figmqom}, 
$M$ and $\mu Q$ are compared for $\Lambda<0$.
The maximal values of $M$ and $\mu Q$
always satisfy $b>0$.

We speculate that 
the disappearance of the spiraling behavior 
is due to the combination of the two effects 
which were argued so far.
The first is the existence of the critical central amplitude of
the temporal component of the vector field $f_{0,{\rm crit}}$,
as discussed in Sec. \ref{sec2d}.
As $|\Lambda|$ increases, 
$f_{0,{\rm crit}}$ decreases,
and hence the allowed region of the central amplitude,
$0< f_0\leq  f_{0,{\rm crit}}$, shrinks.
The second is the overall enhancement of the ADM mass and the Noether charge,
due to the stronger self-interaction for $\Lambda>0$,
as seen in Figs. \ref{figmo0} and \ref{figmqo}.

\subsection{Maximal ADM mass and Noether charge}
\label{sec3b}

\begin{figure}[h]
\unitlength=1.1mm
\begin{center}
  \includegraphics[height=5.0cm,angle=0]{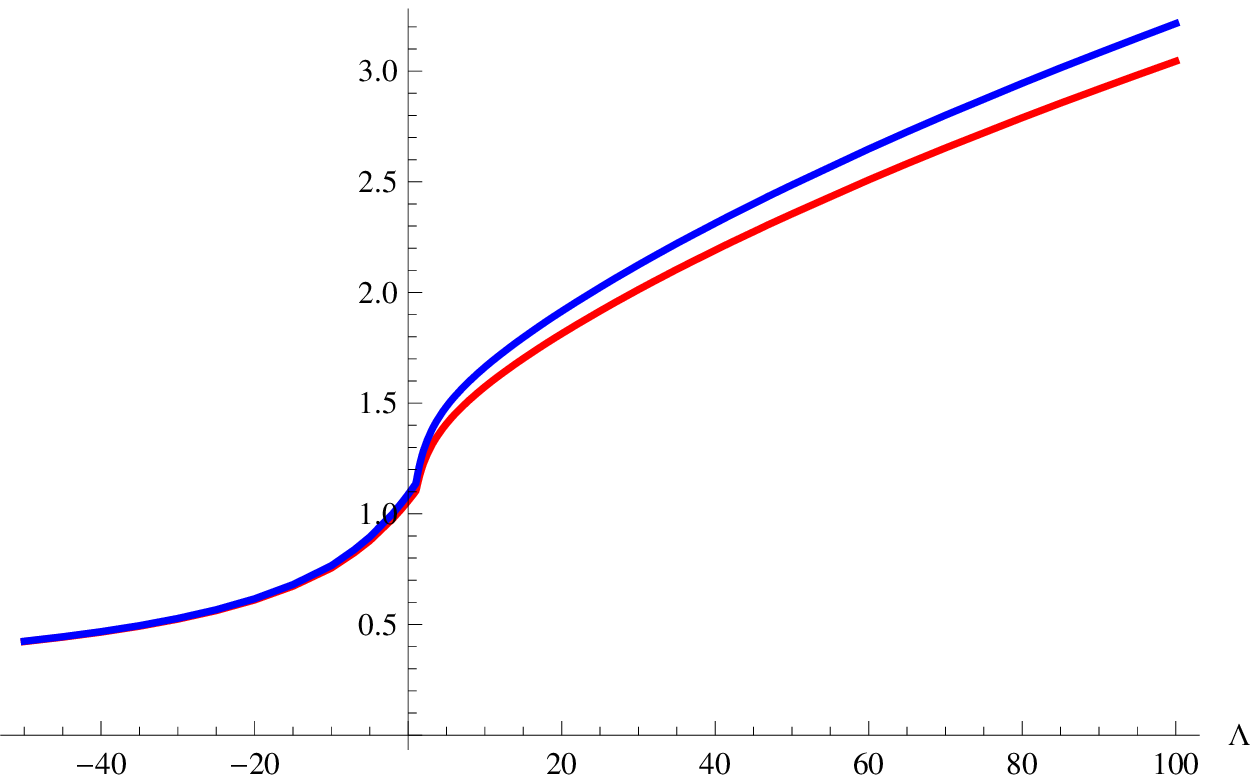}
\vspace{0.3cm}
  \includegraphics[height=5.0cm,angle=0]{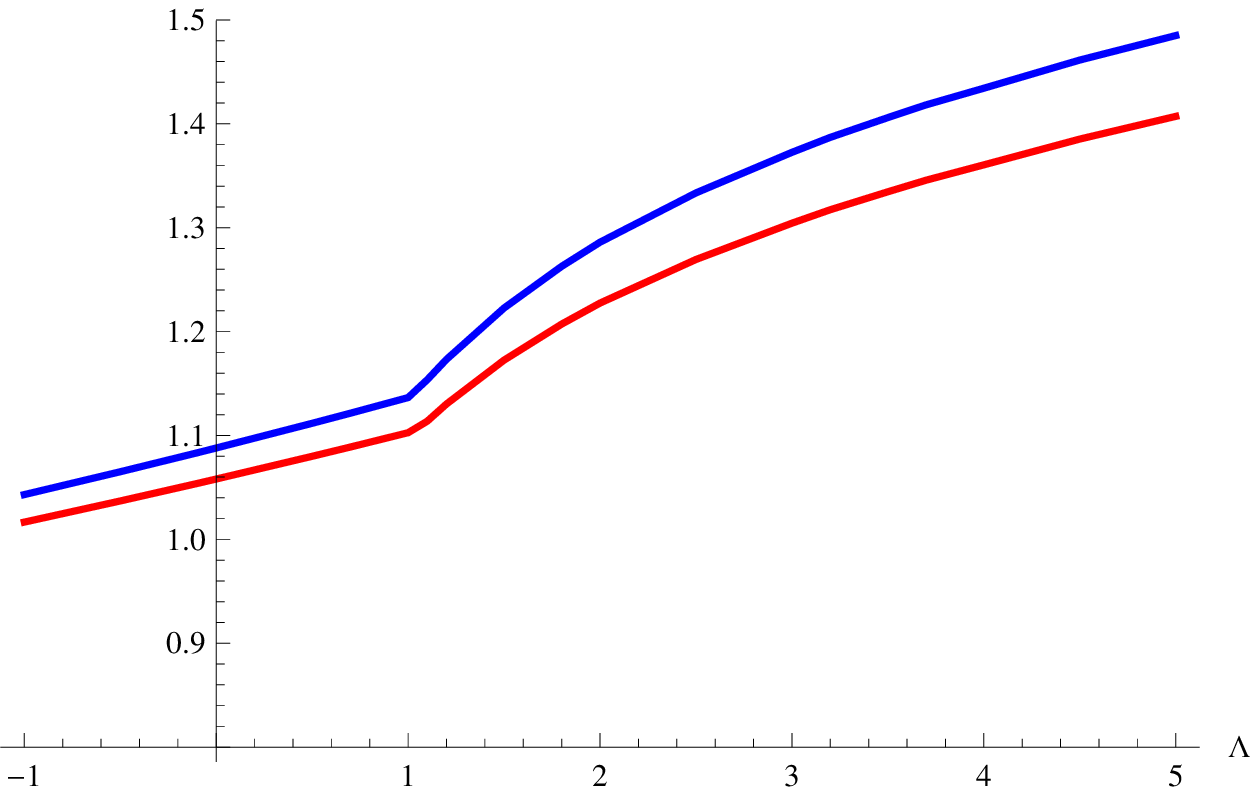}
\caption{
$M_{\rm max}$ and $\mu Q_{\rm max}$
are shown as the functions of $\Lambda$.
The red and blue points correspond 
to $M_{\rm max}$ and $\mu Q_{\rm max}$, respectively.
They are shown in $M_{pl}^2/\mu$.
The lower panel is the enlarged display of the upper one
around $\Lambda=1$.
}
  \label{figmqmax}
\end{center}
\end{figure}

In Fig. \ref{figmqmax},
the maximal values of $M$ and $\mu Q$,
which 
from now on are denoted by $M_{\rm max}$ and $\mu Q_{\rm max}$,
respectively,
are shown as the functions of $\Lambda$.
They are monotonically increasing 
for increasing $\Lambda$.
As already seen in Sec. \ref{sec3a},
$M_{\rm max}$ and $\mu Q_{\rm max}$
obtained for $\Lambda\lesssim 1$
have a different physical origin 
from those obtained for $\Lambda\gtrsim1$,
which 
can be observed as the break around $\Lambda\simeq 1$.
From $\Lambda \gg 1$,
$M_{\rm max}$ and $\mu Q_{\rm max}$
correspond to the values of $M$ and $Q$ 
from the limit of $f_0\to f_{0,\rm crit}$,
respectively.
From the data of $M_{\rm max}$ and $\mu Q_{\rm max}$ 
for $20\leq\Lambda\leq 50$,
the fitting formulas
\begin{subequations}
\label{fitting}
\begin{align}
M_{\rm max}
&\approx 
\frac{M_{pl}^2}{\mu}
\left[
1.383
-0.005099 \frac{\sqrt{\lambda}M_{pl}}{\mu}
\right.
\nonumber\\
&
\left.
+
\left(
-0.03967
+0.005675\frac{\sqrt{\lambda}M_{pl}}{\mu}
\right)
\ln 
\left(\frac{\lambda M_{pl}^2}{\mu^2}\right)
\right],
\\
\mu Q_{\rm max}
&\approx
\frac{M_{pl}^2}{\mu}
\left[
1.500
-0.001989 \frac{\sqrt{\lambda} M_{pl}}{\mu}
\right.
\nonumber\\
&
\left.
+
\left(
-0.05386
+0.005694  \frac{\sqrt{\lambda}M_{pl}}{\mu}
\right)
\ln 
\left(\frac{\lambda M_{pl}^2}{\mu^2}\right)
\right],
\end{align}
\end{subequations}
are obtained, 
respectively,
which can also 
fit the data of $M_{\rm max}$ and $\mu Q_{\rm max}$ for $50<\Lambda\leq 250$
very well.
Thus, 
for a sufficiently large value of $\Lambda\gg 1$,
$M_{\rm max}$ and $\mu Q_{\rm max}$ become of  
${\cal O}[\sqrt{\lambda}M_{pl}^3/\mu^2 \ln (\lambda M_{pl}^2/\mu^2)]$,
which is different from the case of the BS solutions
in the Einstein-scalar theory \eqref{mmax}.
Note that 
for $\Lambda=0$
we recover
\begin{align}
M_{\rm max}\approx 1.058M_{pl}^2/\mu ,
\quad
\mu Q_{\rm max}\approx 1.088M_{pl}^2/\mu,
\end{align} 
obtained in Ref. \cite{Brito:2015pxa}.

\subsection{Binding energy, compactness, and stability}
\label{sec3c}

\begin{figure}[h]
\unitlength=1.1mm
\begin{center}
  \includegraphics[height=5.0cm,angle=0]{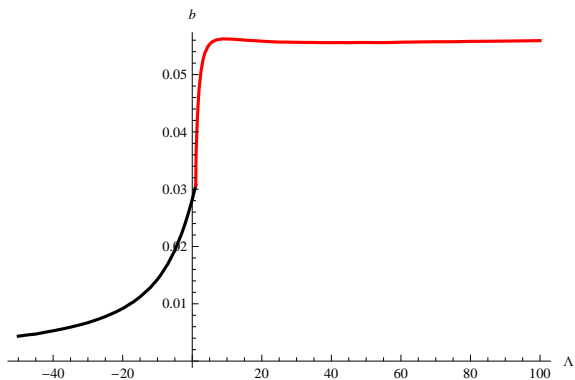}
\caption{
$b_{\rm max}:=\mu Q_{\rm max}/M_{\rm max}-1$
is shown as the function of $\Lambda$.
The black curve corresponds 
to the cases
in which $M_{\rm max}$ and $\mu Q_{\rm max}$
correspond to their local maximal values
obtained at the intermediate value of $f_0< f_{0,{\rm crit}}$,
while 
the red curve 
corresponds to the case 
in which $M_{\rm max}$ and $\mu Q_{\rm max}$
are obtained from the limit of $f_0\to f_{0,{\rm crit}}$.
}
  \label{figbmax}
\end{center}
\end{figure}

In Fig. \ref{figbmax},
the relative binding energy defined in Eq. \eqref{binding}
for the vector BS solutions with $M_{\rm max}$ and $\mu Q_{\rm max}$,
$b_{\rm max}:=\mu Q_{\rm max}/M_{\rm max}-1$,
is shown as the function of $\Lambda$.
The black curve $(\Lambda<1)$ corresponds 
to the cases
in which $M_{\rm max}$ and $\mu Q_{\rm max}$
correspond to their local maximal values
obtained at the intermediate value of $f_0< f_{0,{\rm crit}}$,
while 
the red curve $(\Lambda>1)$
corresponds to the cases 
in which $M_{\rm max}$ and $\mu Q_{\rm max}$
obtained at the intermediate value of $f_0< f_{0,{\rm crit}}$.
Note that for all the values of $\Lambda$, 
$b_{\rm max}>0$,
and hence the vector BS solutions with $M_{\rm max}$ and $\mu Q_{\rm max}$
are always gravitationally bound.
For $\Lambda\lesssim 1$, 
$b_{\rm max}$ is an increasing function of $\Lambda$,
while for $\Lambda\gtrsim 1$
it rapidly increases but eventually 
approaches the constant value $\simeq  0.056$.

\begin{figure}[h]
\unitlength=1.1mm
\begin{center}
  \includegraphics[height=5.0cm,angle=0]{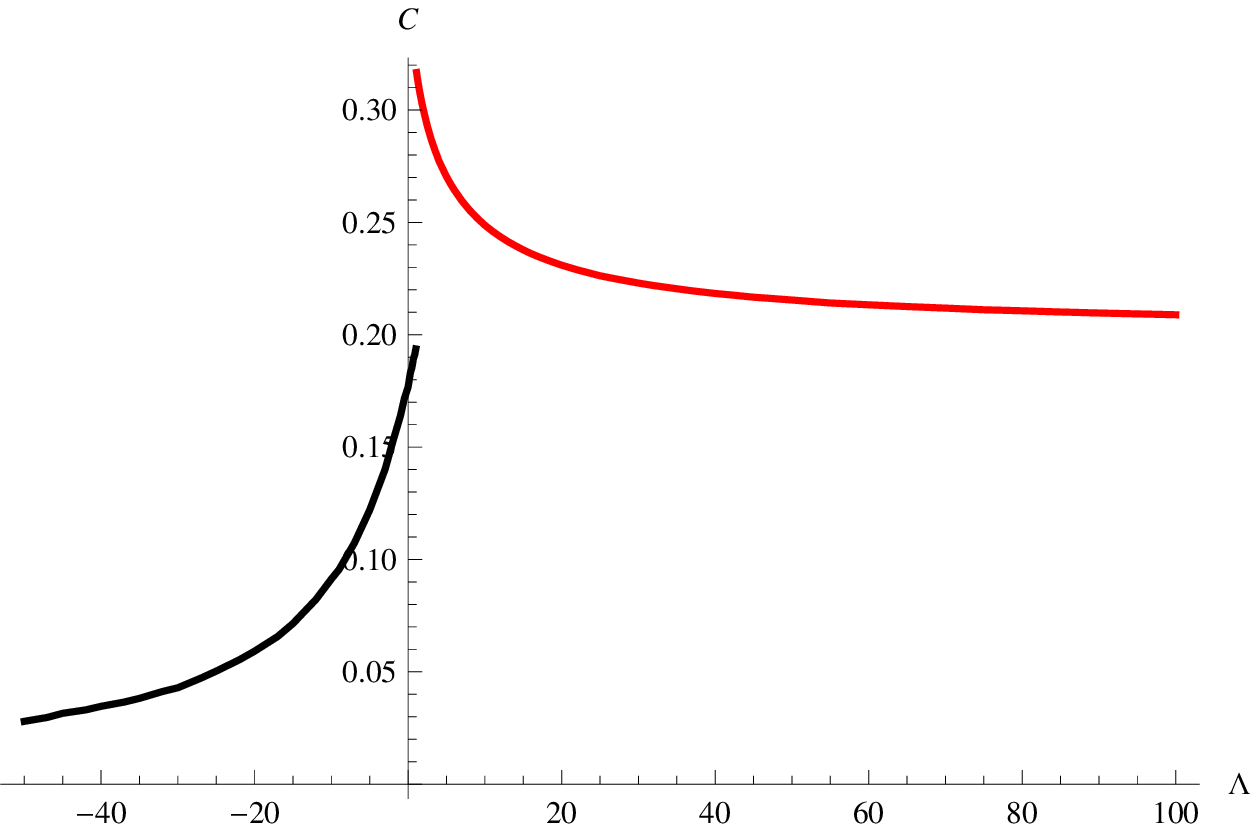}
\caption{
${\cal C}$ for the solutions with $M_{\rm max}$ and $\mu Q_{\rm max}$
is shown as the function of $\Lambda$.
The black curve corresponds 
to the cases
in which $M_{\rm max}$ and $\mu Q_{\rm max}$
correspond to their local maximal values
obtained at the intermediate value of $f_0< f_{0,{\rm crit}}$,
while 
the red curve
corresponds to the case 
in which $M_{\rm max}$ and $\mu Q_{\rm max}$
are obtained from the limit of $f_0\to f_{0,{\rm crit}}$.
}
  \label{figcmax}
\end{center}
\end{figure}

In Fig. \ref{figcmax},
the compactness ${\cal C}$ defined in Eq. \eqref{compactness} 
for the vector BS solutions with $M_{\rm max}$ and $\mu Q_{\rm max}$
is shown as the function of $\Lambda$.
The black curve corresponds 
to the cases
in which $M_{\rm max}$ and $\mu Q_{\rm max}$
correspond to their local maximal values
obtained at the intermediate value of $f_0<f_{0,{\rm crit}}$,
while 
the red curve
corresponds to the case 
in which $M_{\rm max}$ and $\mu Q_{\rm max}$
are obtained from the limit of $f_0\to f_{0,{\rm crit}}$.
The clear discontinuity on the value of ${\cal C}$
 exists around $\Lambda=1$.
For $\Lambda\gtrsim 1$,
${\cal C}$ is always larger than $0.2$,
but cannot exceed $0.32$.
Since photon spheres 
could be formed for a spherically symmetric compact object 
whose compactness is greater than $1/3=0.333\cdots$,
no photon spheres would be formed
around the vector BSs. 
For $\Lambda<1$ including negative values,
${\cal C}$ is less than $0.2$
and gradually decreases
as $|\Lambda|$ increases. 
Note that the compactness \eqref{compactness} 
was defined with the effective radius 
${\cal R}$ defined in \eqref{radius}
outside which the vector field $A_\mu$ does not completely vanish
and the spacetime geometry is not precisely given by the 
vacuum Schwarzschild solution.

\begin{figure}[h]
\unitlength=1.1mm
\begin{center}
  \includegraphics[height=5.0cm,angle=0]{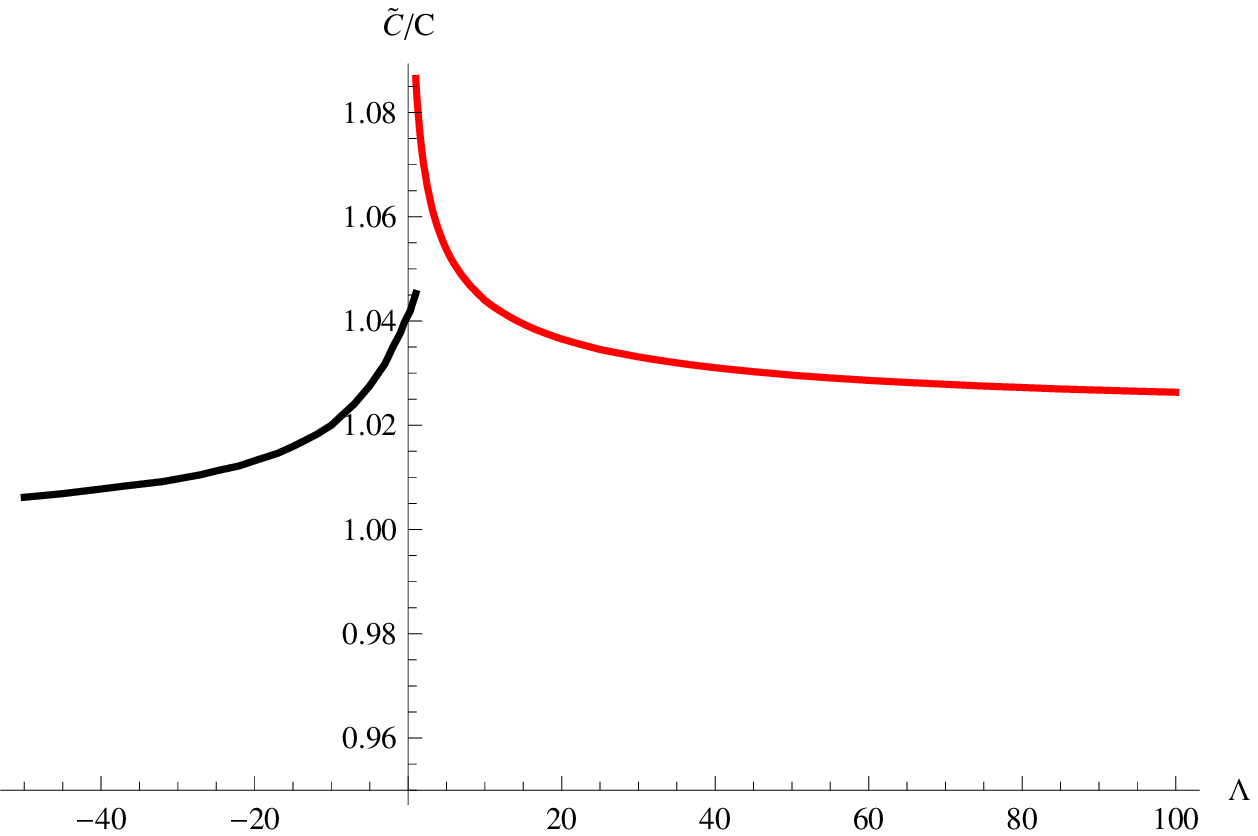}
\caption{
$\tilde{\cal C}/{\cal C}[={\cal R}/\tilde{\cal R}]$ 
for the solutions with $M_{\rm max}$ and $\mu Q_{\rm max}$
is shown as the function of $\Lambda$.
The black curve corresponds 
to the cases
in which $M_{\rm max}$ and $\mu Q_{\rm max}$
correspond to their local maximal values
obtained at the intermediate value of $f_0< f_{0,{\rm crit}}$,
while 
the red curve
corresponds to the case 
in which $M_{\rm max}$ and $\mu Q_{\rm max}$
are obtained from the limit of $f_0\to f_{0,{\rm crit}}$.
}
  \label{figcrmax}
\end{center}
\end{figure}

In order to see 
how the effective compactness depends 
on the definition of the effective radius,
${\cal C}$ defined in Eq. \eqref{compactness}
is compared
with another definition of the effective compactness,
for example,
\begin{align}
\label{compactness2}
\tilde{\cal C}:=\frac{GM}{\tilde \cR}
    =\frac{m_\infty}{\tilde \cR},
\end{align}
for another definition of the effective radius
\begin{align}
\tilde\cR
:=
\left(
\int_0^\infty dr\, r^3 \rho^{(A)}
\right)
\Big{/}
\left(
\int_0^\infty dr\, r^2 \rho^{(A)}
\right),
\end{align}
where
$\rho^{(A)}:=-T^{(A){\hat t}}{}_{\hat t}$ [see Eq. \eqref{tf}] 
is the energy density of the vector field \cite{Schunck:2003kk}.
In Fig. \ref{figcrmax},
the ratio $\tilde{\cal C}/{\cal C}[={\cal R}/\tilde{\cal R}]$
for the solutions with $M_{\rm max}$ and $\mu Q_{\rm max}$
is shown 
as the function of $\Lambda$.
The black curve $(\Lambda<1)$ corresponds 
to the cases
in which $M_{\rm max}$ and $\mu Q_{\rm max}$
correspond to their local maximal values
obtained at the intermediate value of $f_0< f_{0,{\rm crit}}$,
while 
the red curve $(\Lambda>1)$
corresponds to the cases 
in which $M_{\rm max}$ and $\mu Q_{\rm max}$
obtained at the intermediate value of $f_0< f_{0,{\rm crit}}$.
We find 
that for all values of $\Lambda$,
$\tilde{\cal C}/{\cal C}>1$ and hence $\tilde{\cal R}/{\cal R}<1$,
but the deviation from unity is at most 9\%, namely, 
$(\tilde{\cal C}/{\cal C})_{\rm max}-1<0.09$.
The maximal deviation from unity
arises for $\Lambda\simeq 1$,
where $\tilde{\cal C}/{\cal C}\simeq1.087$.
As $|\Lambda|$ increases,
$\tilde{\cal C}/{\cal C}$ decreases toward unity. 
Thus, 
in most cases
the difference between ${\cal C}$ and $\tilde{\cal C}$
is not so quantitatively significant.
But for the solutions with $\Lambda\simeq 1$,
$\tilde{\cal C}$ takes the maximal value
$\tilde{\cal C}> 0.34$ which exceeds $1/3$,
while ${\cal C}<0.32$.
Therefore,
the ambiguity in the definition of 
the effective compactness
around $\Lambda\simeq 1$
makes it unclear
whether 
photon spheres
can be formed 
around the most compact vector BSs
in the presence of the quartic order self-interaction,
or not. 
For the more precise comparison with the other compact objects
such as the BHs or NSs
more careful analyses are requested,
and the possible formation of the photon spheres
should be judged by explicitly analyzing null geodesics 
around the vector BSs,
which will be left for a future study. 

As we mentioned in Sec. \ref{sec1},
for $\Lambda=0$ 
the vector BS solution with $M_{\rm max}$ and $\mu Q_{\rm max}$
corresponds to the critical solution 
which divides the stable and unstable vector BS solutions \cite{Brito:2015pxa},
as in the case of the scalar BS solutions \cite{Gleiser:1988ih,Hawley:2000dt}.
As mentioned previously, 
for all the values of $\Lambda$
the vector BS solutions with $M_{\rm max}$ and $\mu Q_{\rm max}$
satisfy $M_{\rm max}<\mu Q_{\rm max}$,
and they are always gravitationally bound.
For $\Lambda\lesssim 1$
the solutions with $M_{\rm max}$ and $\mu Q_{\rm max}$
obtained from their local maximal values
are also expected to be dynamically stable,
and for $\Lambda\gtrsim 1$
those with $M_{\rm max}$ and $\mu Q_{\rm max}$
obtained from the limit of $f_0\to f_{0,{\rm crit}}$
would also be dynamically stable.
Thus,
for all values of $\Lambda$
the vector BS solution with $M_{\rm max}$ and $\mu Q_{\rm max}$
is expected to be stable.

\section{Conclusion}
\label{sec4}

In this paper,
we have investigated the BS solutions
in the Einstein-Proca theory with the quartic order self-interaction 
as well as the mass \eqref{action}.
While the properties of the BS solutions
in the Einstein-Proca theory with the mass $\mu^2 \bar{A}^\mu A_\mu/2$
have a lot of similarities
with those of the BS solutions 
in the Einstein-scalar theory with the mass $\mu_\phi^2|\phi|^2/2$,
we have found that 
once the quartic order self-interaction $\lambda (\bar{A}^\mu A_\mu)^2/4$
is included into the action,
the properties of the vector BS solutions 
become very distinct
from those of the scalar BS solutions 
with the quartic order self-interaction $\lambda_\phi |\phi|^4/4$.

First,
we have formulated
the basic equations 
to find the BS solutions in the Einstein-Proca theory.
Assuming the static and spherically symmetric metric ansatz \eqref{metric_ansatz}
and the monochromatic oscillation of the vector field in time \eqref{vector_ansatz},
the EOM could be rewritten 
into a set of the evolution equations in the radial direction
\eqref{ms_evolv} and \eqref{a0a1_evolv2}.
Then, the boundary conditions for the metric and vector field variables
were derived by solving the evolution equations 
in the vicinity of the center.
For the frequencies chosen to be the eigenvalues of the BS solutions,
Eqs. \eqref{ms_evolv} and \eqref{a0a1_evolv2}
were able to be numerically integrated,
and 
the metric exponentially approaches the Schwarzschild form \eqref{metric},
while the two components of the vector field exponentially approach $0$.

The clear difference between the cases of the scalar and vector fields
appearing
in the presence of the quartic order self-interaction
was
that in the case of the Einstein-Proca theory 
there is the critical amplitude of the temporal component 
of the vector field at the center,
above which no vector BS solution could be obtained.
Moreover, 
it was found that the qualitative behavior of the vector BSs was different across $\Lambda\simeq 1$,
where $\Lambda$ is the dimensionless coupling constant
defined in Eq. \eqref{Lambda}. 
For $\Lambda\lesssim 1$, 
including the negative values of $\Lambda$,
the behavior of the BS solutions was very similar to the case of $\Lambda=0$.
In this case, 
the maximal values of the ADM mass and Noether charge
correspond to their local maximal values
obtained at the intermediate value of $f_0<f_{0,{\rm crit}}$,
and the compactness defined in Eq. \eqref{compactness} could not exceed $0.20$.
On the other hand, 
for $\Lambda> 1$, 
they
could be obtained from the critical central amplitude
of the temporal component of the vector field,
and 
the compactness was always greater than $0.20$
but could not exceed $1/3=0.333\cdots$,
below which photon spheres would be absent.
However, 
for the most compact vector BS solutions obtained for $\Lambda\simeq 1$
the ratio of the two different definitions of the effective compactness
\eqref{compactness} and \eqref{compactness2}
was close to $1.09$,
and it is still unclear 
whether the photon spheres could be formed around them or not,
which will require further studies.
For $\Lambda\gg 1$
the maximal values of the ADM mass and Noether charge
could be fitted by the formulas \eqref{fitting},
which were of ${\cal O}[\sqrt{\lambda}M_{pl}^3/\mu^2 \ln (\lambda M_{pl}^2/\mu^2)]$,
and slightly larger than the Chandrasekhar mass for the fermions with the same mass $\mu$.

There are a lot of remaining issues,
e.g., 
the stability analysis against the radial and nonradial perturbations,
the implications for the future gravitational wave observations,
and
the BS solutions 
in more general class of the complex vector-tensor theories.
They will be left for future work.

\section*{Acknowledgements}
This work was supported by FCT-Portugal through Grant No. SFRH/BPD/88299/2012.
We thank
the Yukawa Institute for Theoretical Physics, Kyoto University
for the hospitality
during the workshop ``Gravity and Cosmology 2018 '' (YITP-T-17-02)
and the  symposium ``General Relativity $-$ The Next Generation --'' (YKIS2018a).

\bibliography{bibmonster} 

\begin{thebibliography}{25}%
\makeatletter
\providecommand \@ifxundefined [1]{%
 \@ifx{#1\undefined}
}%
\providecommand \@ifnum [1]{%
 \ifnum #1\expandafter \@firstoftwo
 \else \expandafter \@secondoftwo
 \fi
}%
\providecommand \@ifx [1]{%
 \ifx #1\expandafter \@firstoftwo
 \else \expandafter \@secondoftwo
 \fi
}%
\providecommand \natexlab [1]{#1}%
\providecommand \enquote  [1]{``#1''}%
\providecommand \bibnamefont  [1]{#1}%
\providecommand \bibfnamefont [1]{#1}%
\providecommand \citenamefont [1]{#1}%
\providecommand \href@noop [0]{\@secondoftwo}%
\providecommand \href [0]{\begingroup \@sanitize@url \@href}%
\providecommand \@href[1]{\@@startlink{#1}\@@href}%
\providecommand \@@href[1]{\endgroup#1\@@endlink}%
\providecommand \@sanitize@url [0]{\catcode `\\12\catcode `\$12\catcode
  `\&12\catcode `\#12\catcode `\^12\catcode `\_12\catcode `\%12\relax}%
\providecommand \@@startlink[1]{}%
\providecommand \@@endlink[0]{}%
\providecommand \url  [0]{\begingroup\@sanitize@url \@url }%
\providecommand \@url [1]{\endgroup\@href {#1}{\urlprefix }}%
\providecommand \urlprefix  [0]{URL }%
\providecommand \Eprint [0]{\href }%
\providecommand \doibase [0]{http://dx.doi.org/}%
\providecommand \selectlanguage [0]{\@gobble}%
\providecommand \bibinfo  [0]{\@secondoftwo}%
\providecommand \bibfield  [0]{\@secondoftwo}%
\providecommand \translation [1]{[#1]}%
\providecommand \BibitemOpen [0]{}%
\providecommand \bibitemStop [0]{}%
\providecommand \bibitemNoStop [0]{.\EOS\space}%
\providecommand \EOS [0]{\spacefactor3000\relax}%
\providecommand \BibitemShut  [1]{\csname bibitem#1\endcsname}%
\let\auto@bib@innerbib\@empty
\bibitem [{\citenamefont {Abbott}\ \emph
  {et~al.}(2016{\natexlab{a}})\citenamefont {Abbott} \emph
  {et~al.}}]{Abbott:2016blz}%
  \BibitemOpen
  \bibfield  {author} {\bibinfo {author} {\bibfnamefont {B.~ P}\ \bibnamefont
  {Abbott}} \emph {et~al.} (\bibinfo {collaboration} {Virgo, LIGO
  Scientific}),\ }\bibfield  {title} {\enquote {\bibinfo {title} {{Observation
  of Gravitational Waves from a Binary Black Hole Merger}},}\ }\href {\doibase
  10.1103/PhysRevLett.116.061102} {\bibfield  {journal} {\bibinfo  {journal}
  {Phys. Rev. Lett.}\ }\textbf {\bibinfo {volume} {116}},\ \bibinfo {pages}
  {061102} (\bibinfo {year} {2016}{\natexlab{a}})},\ \Eprint
  {http://arxiv.org/abs/1602.03837} {arXiv:1602.03837 [gr-qc]} \BibitemShut
  {NoStop}%
\bibitem [{\citenamefont {Abbott}\ \emph
  {et~al.}(2016{\natexlab{b}})\citenamefont {Abbott} \emph
  {et~al.}}]{Abbott:2016nmj}%
  \BibitemOpen
  \bibfield  {author} {\bibinfo {author} {\bibfnamefont {B.~P.}\ \bibnamefont
  {Abbott}} \emph {et~al.} (\bibinfo {collaboration} {Virgo, LIGO
  Scientific}),\ }\bibfield  {title} {\enquote {\bibinfo {title} {{GW151226:
  Observation of Gravitational Waves from a 22-Solar-Mass Binary Black Hole
  Coalescence}},}\ }\href {\doibase 10.1103/PhysRevLett.116.241103} {\bibfield
  {journal} {\bibinfo  {journal} {Phys. Rev. Lett.}\ }\textbf {\bibinfo
  {volume} {116}},\ \bibinfo {pages} {241103} (\bibinfo {year}
  {2016}{\natexlab{b}})},\ \Eprint {http://arxiv.org/abs/1606.04855}
  {arXiv:1606.04855 [gr-qc]} \BibitemShut {NoStop}%
\bibitem [{\citenamefont {Abbott}\ \emph {et~al.}(2017)\citenamefont {Abbott}
  \emph {et~al.}}]{TheLIGOScientific:2017qsa}%
  \BibitemOpen
  \bibfield  {author} {\bibinfo {author} {\bibfnamefont {B.~P.}\ \bibnamefont
  {Abbott}} \emph {et~al.} (\bibinfo {collaboration} {Virgo, LIGO
  Scientific}),\ }\bibfield  {title} {\enquote {\bibinfo {title} {{GW170817:
  Observation of Gravitational Waves from a Binary Neutron Star Inspiral}},}\
  }\href {\doibase 10.1103/PhysRevLett.119.161101} {\bibfield  {journal}
  {\bibinfo  {journal} {Phys. Rev. Lett.}\ }\textbf {\bibinfo {volume} {119}},\
  \bibinfo {pages} {161101} (\bibinfo {year} {2017})},\ \Eprint
  {http://arxiv.org/abs/1710.05832} {arXiv:1710.05832 [gr-qc]} \BibitemShut
  {NoStop}%
\bibitem [{\citenamefont {Berti}\ \emph {et~al.}(2015)\citenamefont {Berti}
  \emph {et~al.}}]{Berti:2015itd}%
  \BibitemOpen
  \bibfield  {author} {\bibinfo {author} {\bibfnamefont {Emanuele}\
  \bibnamefont {Berti}} \emph {et~al.},\ }\bibfield  {title} {\enquote
  {\bibinfo {title} {{Testing General Relativity with Present and Future
  Astrophysical Observations}},}\ }\href {\doibase
  10.1088/0264-9381/32/24/243001} {\bibfield  {journal} {\bibinfo  {journal}
  {Class. Quant. Grav.}\ }\textbf {\bibinfo {volume} {32}},\ \bibinfo {pages}
  {243001} (\bibinfo {year} {2015})},\ \Eprint
  {http://arxiv.org/abs/1501.07274} {arXiv:1501.07274 [gr-qc]} \BibitemShut
  {NoStop}%
\bibitem [{\citenamefont {Herdeiro}\ and\ \citenamefont
  {Radu}(2015)}]{Herdeiro:2015waa}%
  \BibitemOpen
  \bibfield  {author} {\bibinfo {author} {\bibfnamefont {Carlos A.~R.}\
  \bibnamefont {Herdeiro}}\ and\ \bibinfo {author} {\bibfnamefont {Eugen}\
  \bibnamefont {Radu}},\ }\bibfield  {title} {\enquote {\bibinfo {title}
  {{Asymptotically flat black holes with scalar hair: a review}},}\ }\bibfield
  {booktitle} {\emph {\bibinfo {booktitle} {{Proceedings, 7th Black Holes
  Workshop 2014}}},\ }\href {\doibase 10.1142/S0218271815420146} {\bibfield
  {journal} {\bibinfo  {journal} {Int. J. Mod. Phys.}\ }\textbf {\bibinfo
  {volume} {D24}},\ \bibinfo {pages} {1542014} (\bibinfo {year} {2015})},\
  \Eprint {http://arxiv.org/abs/1504.08209} {arXiv:1504.08209 [gr-qc]}
  \BibitemShut {NoStop}%
\bibitem [{\citenamefont {Doneva}\ and\ \citenamefont
  {Pappas}(2017)}]{Doneva:2017jop}%
  \BibitemOpen
  \bibfield  {author} {\bibinfo {author} {\bibfnamefont {Daniela~D.}\
  \bibnamefont {Doneva}}\ and\ \bibinfo {author} {\bibfnamefont {George}\
  \bibnamefont {Pappas}},\ }\bibfield  {title} {\enquote {\bibinfo {title}
  {{Universal Relations and Alternative Gravity Theories}},}\ }\href@noop {} {\
   (\bibinfo {year} {2017})},\ \Eprint {http://arxiv.org/abs/1709.08046}
  {arXiv:1709.08046 [gr-qc]} \BibitemShut {NoStop}%
\bibitem [{\citenamefont {Berti}\ \emph
  {et~al.}(2018{\natexlab{a}})\citenamefont {Berti}, \citenamefont {Yagi},\
  and\ \citenamefont {Yunes}}]{Berti:2018cxi}%
  \BibitemOpen
  \bibfield  {author} {\bibinfo {author} {\bibfnamefont {Emanuele}\
  \bibnamefont {Berti}}, \bibinfo {author} {\bibfnamefont {Kent}\ \bibnamefont
  {Yagi}}, \ and\ \bibinfo {author} {\bibfnamefont {Nicolas}\ \bibnamefont
  {Yunes}},\ }\bibfield  {title} {\enquote {\bibinfo {title} {{Extreme Gravity
  Tests with Gravitational Waves from Compact Binary Coalescences: (I)
  Inspiral-Merger}},}\ }\href {\doibase 10.1007/s10714-018-2362-8} {\bibfield
  {journal} {\bibinfo  {journal} {Gen. Rel. Grav.}\ }\textbf {\bibinfo {volume}
  {50}},\ \bibinfo {pages} {46} (\bibinfo {year} {2018}{\natexlab{a}})},\
  \Eprint {http://arxiv.org/abs/1801.03208} {arXiv:1801.03208 [gr-qc]}
  \BibitemShut {NoStop}%
\bibitem [{\citenamefont {Berti}\ \emph
  {et~al.}(2018{\natexlab{b}})\citenamefont {Berti}, \citenamefont {Yagi},
  \citenamefont {Yang},\ and\ \citenamefont {Yunes}}]{Berti:2018vdi}%
  \BibitemOpen
  \bibfield  {author} {\bibinfo {author} {\bibfnamefont {Emanuele}\
  \bibnamefont {Berti}}, \bibinfo {author} {\bibfnamefont {Kent}\ \bibnamefont
  {Yagi}}, \bibinfo {author} {\bibfnamefont {Huan}\ \bibnamefont {Yang}}, \
  and\ \bibinfo {author} {\bibfnamefont {Nicolas}\ \bibnamefont {Yunes}},\
  }\bibfield  {title} {\enquote {\bibinfo {title} {{Extreme Gravity Tests with
  Gravitational Waves from Compact Binary Coalescences: (II) Ringdown}},}\
  }\href {\doibase 10.1007/s10714-018-2372-6} {\bibfield  {journal} {\bibinfo
  {journal} {Gen. Rel. Grav.}\ }\textbf {\bibinfo {volume} {50}},\ \bibinfo
  {pages} {49} (\bibinfo {year} {2018}{\natexlab{b}})},\ \Eprint
  {http://arxiv.org/abs/1801.03587} {arXiv:1801.03587 [gr-qc]} \BibitemShut
  {NoStop}%
\bibitem [{\citenamefont {Cardoso}\ and\ \citenamefont
  {Pani}(2017)}]{Cardoso:2017cqb}%
  \BibitemOpen
  \bibfield  {author} {\bibinfo {author} {\bibfnamefont {Vitor}\ \bibnamefont
  {Cardoso}}\ and\ \bibinfo {author} {\bibfnamefont {Paolo}\ \bibnamefont
  {Pani}},\ }\bibfield  {title} {\enquote {\bibinfo {title} {{Tests for the
  existence of black holes through gravitational wave echoes}},}\ }\href
  {\doibase 10.1038/s41550-017-0225-y} {\bibfield  {journal} {\bibinfo
  {journal} {Nat. Astron.}\ }\textbf {\bibinfo {volume} {1}},\ \bibinfo {pages}
  {586--591} (\bibinfo {year} {2017})},\ \Eprint
  {http://arxiv.org/abs/1709.01525} {arXiv:1709.01525 [gr-qc]} \BibitemShut
  {NoStop}%
\bibitem [{\citenamefont {Kaup}(1968)}]{Kaup:1968zz}%
  \BibitemOpen
  \bibfield  {author} {\bibinfo {author} {\bibfnamefont {David~J.}\
  \bibnamefont {Kaup}},\ }\bibfield  {title} {\enquote {\bibinfo {title}
  {{Klein-Gordon Geon}},}\ }\href {\doibase 10.1103/PhysRev.172.1331}
  {\bibfield  {journal} {\bibinfo  {journal} {Phys. Rev.}\ }\textbf {\bibinfo
  {volume} {172}},\ \bibinfo {pages} {1331--1342} (\bibinfo {year}
  {1968})}\BibitemShut {NoStop}%
\bibitem [{\citenamefont {Ruffini}\ and\ \citenamefont
  {Bonazzola}(1969)}]{Ruffini:1969qy}%
  \BibitemOpen
  \bibfield  {author} {\bibinfo {author} {\bibfnamefont {Remo}\ \bibnamefont
  {Ruffini}}\ and\ \bibinfo {author} {\bibfnamefont {Silvano}\ \bibnamefont
  {Bonazzola}},\ }\bibfield  {title} {\enquote {\bibinfo {title} {{Systems of
  selfgravitating particles in general relativity and the concept of an
  equation of state}},}\ }\href {\doibase 10.1103/PhysRev.187.1767} {\bibfield
  {journal} {\bibinfo  {journal} {Phys. Rev.}\ }\textbf {\bibinfo {volume}
  {187}},\ \bibinfo {pages} {1767--1783} (\bibinfo {year} {1969})}\BibitemShut
  {NoStop}%
\bibitem [{\citenamefont {Friedberg}\ \emph {et~al.}(1987)\citenamefont
  {Friedberg}, \citenamefont {Lee},\ and\ \citenamefont
  {Pang}}]{Friedberg:1986tp}%
  \BibitemOpen
  \bibfield  {author} {\bibinfo {author} {\bibfnamefont {R.}~\bibnamefont
  {Friedberg}}, \bibinfo {author} {\bibfnamefont {T.~D.}\ \bibnamefont {Lee}},
  \ and\ \bibinfo {author} {\bibfnamefont {Y.}~\bibnamefont {Pang}},\
  }\bibfield  {title} {\enquote {\bibinfo {title} {{MINI - SOLITON STARS}},}\
  }\href {\doibase 10.1103/PhysRevD.35.3640} {\bibfield  {journal} {\bibinfo
  {journal} {Phys. Rev.}\ }\textbf {\bibinfo {volume} {D35}},\ \bibinfo {pages}
  {3640} (\bibinfo {year} {1987})}\BibitemShut {NoStop}%
\bibitem [{\citenamefont {Jetzer}(1992)}]{Jetzer:1991jr}%
  \BibitemOpen
  \bibfield  {author} {\bibinfo {author} {\bibfnamefont {Philippe}\
  \bibnamefont {Jetzer}},\ }\bibfield  {title} {\enquote {\bibinfo {title}
  {{Boson stars}},}\ }\href {\doibase 10.1016/0370-1573(92)90123-H} {\bibfield
  {journal} {\bibinfo  {journal} {Phys. Rept.}\ }\textbf {\bibinfo {volume}
  {220}},\ \bibinfo {pages} {163--227} (\bibinfo {year} {1992})}\BibitemShut
  {NoStop}%
\bibitem [{\citenamefont {Gleiser}(1988)}]{Gleiser:1988rq}%
  \BibitemOpen
  \bibfield  {author} {\bibinfo {author} {\bibfnamefont {Marcelo}\ \bibnamefont
  {Gleiser}},\ }\bibfield  {title} {\enquote {\bibinfo {title} {{Stability of
  Boson Stars}},}\ }\href {\doibase 10.1103/PhysRevD.38.2376} {\bibfield
  {journal} {\bibinfo  {journal} {Phys. Rev.}\ }\textbf {\bibinfo {volume}
  {D38}},\ \bibinfo {pages} {2376} (\bibinfo {year} {1988})}\BibitemShut
  {NoStop}%
\bibitem [{\citenamefont {Lee}\ and\ \citenamefont {Pang}(1989)}]{Lee:1988av}%
  \BibitemOpen
  \bibfield  {author} {\bibinfo {author} {\bibfnamefont {T.~D.}\ \bibnamefont
  {Lee}}\ and\ \bibinfo {author} {\bibfnamefont {Yang}\ \bibnamefont {Pang}},\
  }\bibfield  {title} {\enquote {\bibinfo {title} {{Stability of Mini - Boson
  Stars}},}\ }\href {\doibase 10.1016/0550-3213(89)90365-9} {\bibfield
  {journal} {\bibinfo  {journal} {Nucl. Phys.}\ }\textbf {\bibinfo {volume}
  {B315}},\ \bibinfo {pages} {477} (\bibinfo {year} {1989})}\BibitemShut
  {NoStop}%
\bibitem [{\citenamefont {Gleiser}\ and\ \citenamefont
  {Watkins}(1989)}]{Gleiser:1988ih}%
  \BibitemOpen
  \bibfield  {author} {\bibinfo {author} {\bibfnamefont {Marcelo}\ \bibnamefont
  {Gleiser}}\ and\ \bibinfo {author} {\bibfnamefont {Richard}\ \bibnamefont
  {Watkins}},\ }\bibfield  {title} {\enquote {\bibinfo {title} {{Gravitational
  Stability of Scalar Matter}},}\ }\href {\doibase
  10.1016/0550-3213(89)90627-5} {\bibfield  {journal} {\bibinfo  {journal}
  {Nucl. Phys.}\ }\textbf {\bibinfo {volume} {B319}},\ \bibinfo {pages}
  {733--746} (\bibinfo {year} {1989})}\BibitemShut {NoStop}%
\bibitem [{\citenamefont {Hawley}\ and\ \citenamefont
  {Choptuik}(2000)}]{Hawley:2000dt}%
  \BibitemOpen
  \bibfield  {author} {\bibinfo {author} {\bibfnamefont {Scott~H.}\
  \bibnamefont {Hawley}}\ and\ \bibinfo {author} {\bibfnamefont {Matthew~W.}\
  \bibnamefont {Choptuik}},\ }\bibfield  {title} {\enquote {\bibinfo {title}
  {{Boson stars driven to the brink of black hole formation}},}\ }\href
  {\doibase 10.1103/PhysRevD.62.104024} {\bibfield  {journal} {\bibinfo
  {journal} {Phys. Rev.}\ }\textbf {\bibinfo {volume} {D62}},\ \bibinfo {pages}
  {104024} (\bibinfo {year} {2000})},\ \Eprint
  {http://arxiv.org/abs/gr-qc/0007039} {arXiv:gr-qc/0007039 [gr-qc]}
  \BibitemShut {NoStop}%
\bibitem [{\citenamefont {Sennett}\ \emph {et~al.}(2017)\citenamefont
  {Sennett}, \citenamefont {Hinderer}, \citenamefont {Steinhoff}, \citenamefont
  {Buonanno},\ and\ \citenamefont {Ossokine}}]{Sennett:2017etc}%
  \BibitemOpen
  \bibfield  {author} {\bibinfo {author} {\bibfnamefont {Noah}\ \bibnamefont
  {Sennett}}, \bibinfo {author} {\bibfnamefont {Tanja}\ \bibnamefont
  {Hinderer}}, \bibinfo {author} {\bibfnamefont {Jan}\ \bibnamefont
  {Steinhoff}}, \bibinfo {author} {\bibfnamefont {Alessandra}\ \bibnamefont
  {Buonanno}}, \ and\ \bibinfo {author} {\bibfnamefont {Serguei}\ \bibnamefont
  {Ossokine}},\ }\bibfield  {title} {\enquote {\bibinfo {title}
  {{Distinguishing Boson Stars from Black Holes and Neutron Stars from Tidal
  Interactions in Inspiraling Binary Systems}},}\ }\href {\doibase
  10.1103/PhysRevD.96.024002} {\bibfield  {journal} {\bibinfo  {journal} {Phys.
  Rev.}\ }\textbf {\bibinfo {volume} {D96}},\ \bibinfo {pages} {024002}
  (\bibinfo {year} {2017})},\ \Eprint {http://arxiv.org/abs/1704.08651}
  {arXiv:1704.08651 [gr-qc]} \BibitemShut {NoStop}%
\bibitem [{\citenamefont {Schunck}\ and\ \citenamefont
  {Mielke}(2003)}]{Schunck:2003kk}%
  \BibitemOpen
  \bibfield  {author} {\bibinfo {author} {\bibfnamefont {F.E.}\ \bibnamefont
  {Schunck}}\ and\ \bibinfo {author} {\bibfnamefont {E.W.}\ \bibnamefont
  {Mielke}},\ }\bibfield  {title} {\enquote {\bibinfo {title} {{General
  relativistic boson stars}},}\ }\href {\doibase 10.1088/0264-9381/20/20/201}
  {\bibfield  {journal} {\bibinfo  {journal} {Class.Quant.Grav.}\ }\textbf
  {\bibinfo {volume} {20}},\ \bibinfo {pages} {R301--R356} (\bibinfo {year}
  {2003})},\ \Eprint {http://arxiv.org/abs/0801.0307} {arXiv:0801.0307
  [astro-ph]} \BibitemShut {NoStop}%
\bibitem [{\citenamefont {Colpi}\ \emph {et~al.}(1986)\citenamefont {Colpi},
  \citenamefont {Shapiro},\ and\ \citenamefont {Wasserman}}]{Colpi:1986ye}%
  \BibitemOpen
  \bibfield  {author} {\bibinfo {author} {\bibfnamefont {M.}~\bibnamefont
  {Colpi}}, \bibinfo {author} {\bibfnamefont {S.~L.}\ \bibnamefont {Shapiro}},
  \ and\ \bibinfo {author} {\bibfnamefont {I.}~\bibnamefont {Wasserman}},\
  }\bibfield  {title} {\enquote {\bibinfo {title} {{Boson Stars: Gravitational
  Equilibria of Selfinteracting Scalar Fields}},}\ }\href {\doibase
  10.1103/PhysRevLett.57.2485} {\bibfield  {journal} {\bibinfo  {journal}
  {Phys. Rev. Lett.}\ }\textbf {\bibinfo {volume} {57}},\ \bibinfo {pages}
  {2485--2488} (\bibinfo {year} {1986})}\BibitemShut {NoStop}%
\bibitem [{\citenamefont {Brito}\ \emph {et~al.}(2016)\citenamefont {Brito},
  \citenamefont {Cardoso}, \citenamefont {Herdeiro},\ and\ \citenamefont
  {Radu}}]{Brito:2015pxa}%
  \BibitemOpen
  \bibfield  {author} {\bibinfo {author} {\bibfnamefont {Richard}\ \bibnamefont
  {Brito}}, \bibinfo {author} {\bibfnamefont {Vitor}\ \bibnamefont {Cardoso}},
  \bibinfo {author} {\bibfnamefont {Carlos A.~R.}\ \bibnamefont {Herdeiro}}, \
  and\ \bibinfo {author} {\bibfnamefont {Eugen}\ \bibnamefont {Radu}},\
  }\bibfield  {title} {\enquote {\bibinfo {title} {{Proca stars: Gravitating
  Bose-Einstein condensates of massive spin 1 particles}},}\ }\href {\doibase
  10.1016/j.physletb.2015.11.051} {\bibfield  {journal} {\bibinfo  {journal}
  {Phys. Lett.}\ }\textbf {\bibinfo {volume} {B752}},\ \bibinfo {pages}
  {291--295} (\bibinfo {year} {2016})},\ \Eprint
  {http://arxiv.org/abs/1508.05395} {arXiv:1508.05395 [gr-qc]} \BibitemShut
  {NoStop}%
\bibitem [{\citenamefont {Landea}\ and\ \citenamefont
  {Garcia}(2016)}]{Garcia:2016ldc}%
  \BibitemOpen
  \bibfield  {author} {\bibinfo {author} {\bibfnamefont {Ignacio~Salazar}\
  \bibnamefont {Landea}}\ and\ \bibinfo {author} {\bibfnamefont {Federico}\
  \bibnamefont {Garcia}},\ }\bibfield  {title} {\enquote {\bibinfo {title}
  {{Charged Proca Stars}},}\ }\href {\doibase 10.1103/PhysRevD.94.104006}
  {\bibfield  {journal} {\bibinfo  {journal} {Phys. Rev.}\ }\textbf {\bibinfo
  {volume} {D94}},\ \bibinfo {pages} {104006} (\bibinfo {year} {2016})},\
  \Eprint {http://arxiv.org/abs/1608.00011} {arXiv:1608.00011 [hep-th]}
  \BibitemShut {NoStop}%
\bibitem [{\citenamefont {Brihaye}\ \emph {et~al.}(2017)\citenamefont
  {Brihaye}, \citenamefont {Delplace},\ and\ \citenamefont
  {Verbin}}]{Brihaye:2017inn}%
  \BibitemOpen
  \bibfield  {author} {\bibinfo {author} {\bibfnamefont {Y.}~\bibnamefont
  {Brihaye}}, \bibinfo {author} {\bibfnamefont {Th.}\ \bibnamefont {Delplace}},
  \ and\ \bibinfo {author} {\bibfnamefont {Y.}~\bibnamefont {Verbin}},\
  }\bibfield  {title} {\enquote {\bibinfo {title} {{Proca Q Balls and their
  Coupling to Gravity}},}\ }\href {\doibase 10.1103/PhysRevD.96.024057}
  {\bibfield  {journal} {\bibinfo  {journal} {Phys. Rev.}\ }\textbf {\bibinfo
  {volume} {D96}},\ \bibinfo {pages} {024057} (\bibinfo {year} {2017})},\
  \Eprint {http://arxiv.org/abs/1704.01648} {arXiv:1704.01648 [gr-qc]}
  \BibitemShut {NoStop}%
\bibitem [{\citenamefont {Loginov}(2015)}]{Loginov:2015rya}%
  \BibitemOpen
  \bibfield  {author} {\bibinfo {author} {\bibfnamefont {A.~{\relax Yu}.}\
  \bibnamefont {Loginov}},\ }\bibfield  {title} {\enquote {\bibinfo {title}
  {{Nontopological solitons in the model of the self-interacting complex vector
  field}},}\ }\href {\doibase 10.1103/PhysRevD.91.105028} {\bibfield  {journal}
  {\bibinfo  {journal} {Phys. Rev.}\ }\textbf {\bibinfo {volume} {D91}},\
  \bibinfo {pages} {105028} (\bibinfo {year} {2015})}\BibitemShut {NoStop}%
\bibitem [{\citenamefont {Brihaye}\ and\ \citenamefont
  {Verbin}(2017)}]{Brihaye:2016pld}%
  \BibitemOpen
  \bibfield  {author} {\bibinfo {author} {\bibfnamefont {Y.}~\bibnamefont
  {Brihaye}}\ and\ \bibinfo {author} {\bibfnamefont {Y.}~\bibnamefont
  {Verbin}},\ }\bibfield  {title} {\enquote {\bibinfo {title} {{Proca Q Tubes
  and their Coupling to Gravity}},}\ }\href {\doibase
  10.1103/PhysRevD.95.044027} {\bibfield  {journal} {\bibinfo  {journal} {Phys.
  Rev.}\ }\textbf {\bibinfo {volume} {D95}},\ \bibinfo {pages} {044027}
  (\bibinfo {year} {2017})},\ \Eprint {http://arxiv.org/abs/1611.01803}
  {arXiv:1611.01803 [gr-qc]} \BibitemShut {NoStop}%
\end{thebibliography}%
\end{document}